\documentclass[fleqn,twoside]{article}
\usepackage{espcrc2}

\usepackage{graphicx,xspace}
\usepackage[figuresright]{rotating}
\newcommand{\dalhad}{\im{\Delta\alpha^{(5)}_{had}}}

\newcommand{\AmS}{{\protect\the\textfont2
  A\kern-.1667em\lower.5ex\hbox{M}\kern-.125emS}}

\newcommand{\beq}{\begin{equation}}
\newcommand{\eeq}{\end{equation}}
\newcommand{\bea}{\begin{eqnarray}}
\newcommand{\eea}{\end{eqnarray}}
\newcommand{\bseq}{\begin{subequations}}
\newcommand{\eseq}{\end{subequations}}
\newcommand{\bsea}{\begin{subeqnarray}}
\newcommand{\esea}{\end{subeqnarray}}
\newcommand{\bit}{\begin{itemize}}
\newcommand{\eit}{\end{itemize}}
\newcommand{\ben}{\begin{enumerate}}
\newcommand{\een}{\end{enumerate}}
\newcommand{\bfig}{\begin{figure}}
\newcommand{\efig}{\end{figure}}
\newcommand{\btab}{\begin{table}}
\newcommand{\etab}{\end{table}}
\newcommand{\im}[1]{\ensuremath{#1}\xspace}       
\newcommand{\imx}[1]{\ensuremath{#1}\xspace}       
\newcommand{\etal}{{\em et al.}}
\newcommand{\eg}{{\em e.g.}}
\newcommand{\ie}{{\em i.e.}}

\newcommand{\eV}{\imx{\mathrm{e\kern -0.1em V}}}
\newcommand{\MeV}{\imx{\mathrm{Me\kern -0.1em V}}}
\newcommand{\GeV}{\imx{\mathrm{Ge\kern -0.1em V}}}
\newcommand{\TeV}{\imx{\mathrm{Te\kern -0.1em V}}}

\newcommand{\IM}{\im{\mathrm{I} \kern -0.15 em \mathrm{m}}}         
\newcommand{\RE}{\im{\mathrm{R} \kern -0.15 em \mathrm{e}}}         

\newcommand{\Pchi}{\im{{\raise5pt\hbox{$\chi$}}}}
\newcommand{\Pe}{\im{\mathrm{e}}}            
\newcommand{\Pm}{\im{\mu}}
\newcommand{\Pt}{\im{\tau}}
\newcommand{\Pn}{\im{\nu}}
\newcommand{\Pl}{\im{\ell}}

\newcommand{\Pc}{\im{\mathrm{c}}}
\newcommand{\Pb}{\im{\mathrm{b}}}
\newcommand{\PT}{\im{\mathrm{t}}}
\newcommand{\Pq}{\im{q}}
\newcommand{\Pf}{\im{f}}
\newcommand{\Phad}{\im{\mathrm{had}}}

\newcommand{\PW}{\im{\mathrm{W}}}             

\newcommand{\PZ}{\im{\mathrm{Z}}}
\newcommand{\PH}{\im{\mathrm{H}}}

\newcommand{\MW}{\im{M_{\PW}}}
\newcommand{\MZ}{\im{M_{\PZ}}}
\newcommand{\MH}{\im{M_{\PH}}}

\newcommand{\MT}{\im{M_{\PT}}}

\newcommand{\G}{\im{\Gamma}}                  
\newcommand{\GZ}{\im{\G_{\PZ}}}               

\newcommand{\GZq}{\im{\G_{\Pq\Pq}}}

\newcommand{\GZhad}{\im{\G_{\Phad}}}

\newcommand{\GW}{\im{\G_{\PW}}}               

\newcommand{\A}{\im{\mathrm{A}}}
\newcommand{\V}{\im{\mathrm{V}}}
\newcommand{\R}{\im{\mathrm{R}}}

\newcommand{\g}{\im{g}}                       

\newcommand{\gal}{\im{\g_{\A\Pl}}}
\newcommand{\gvl}{\im{\g_{\V\Pl}}}

\newcommand{\gaf}{\im{\g_{\A\Pf}}}
\newcommand{\gvf}{\im{\g_{\V\Pf}}}

\newcommand{\gsq}{\im{\g^2}}

\newcommand{\gafsq}{\im{\gsq_{\A\Pf}}}
\newcommand{\gvfsq}{\im{\gsq_{\V\Pf}}}

\newcommand{\Ae}{\im{\A_{\Pe}}}

\newcommand{\Al}{\im{\A_{\Pl}}}
\newcommand{\Ab}{\im{\A_{\Pb}}}
\newcommand{\Ac}{\im{\A_{\Pc}}}
\newcommand{\Aq}{\im{\A_{\Pq}}}
\newcommand{\Af}{\im{\A_{\Pf}}}

\newcommand{\Afb}{\im{\A_{\mathrm{fb}}}}

\newcommand{\Afbzl}{\im{\Afb^{0,\Pl}}}
\newcommand{\Afbzb}{\im{\Afb^{0,\Pb}}}
\newcommand{\Afbzc}{\im{\Afb^{0,\Pc}}}
\newcommand{\Afbzq}{\im{\Afb^{0,\Pq}}}
\newcommand{\Afbzf}{\im{\Afb^{0,\Pf}}}

\newcommand{\Rb}{\im{\R_{\Pb}}}
\newcommand{\Rc}{\im{\R_{\Pc}}}
\newcommand{\Rq}{\im{\R_{\Pq}}}

\newcommand{\swsqeffl}{\sin^2\theta_{\mathrm{eff}}^{\mathrm{lept}}}

\newcommand{\Pee}{\im{\Pe^+\Pe^-}}
\newcommand{\Pmm}{\im{\Pm^+\Pm^-}}
\newcommand{\Ptt}{\im{\Pt^+\Pt^-}}
\newcommand{\Pll}{\im{\Pl^+\Pl^-}}

\newcommand{\Pnn}{\im{\Pn\overline{\Pn}}}
\newcommand{\Pqq}{\im{\Pq\overline{\Pq}}}
\newcommand{\PTT}{\im{\PT\overline{\PT}}}
\newcommand{\Pbb}{\im{\Pb\overline{\Pb}}}

\newcommand{\Pff}{\im{\Pf\overline{\Pf}}}
\newcommand{\PWW}{\im{\PW^+\PW^-}}
\newcommand{\PZZ}{\im{\PZ\PZ}}

\newcommand{\Peeee}{\im{\Pee \kern -0.35em \rightarrow\Pee}}
\newcommand{\Peemm}{\im{\Pee \kern -0.35em \rightarrow\Pmm}}
\newcommand{\Peett}{\im{\Pee \kern -0.35em \rightarrow\Ptt}}
\newcommand{\Peell}{\im{\Pee \kern -0.35em \rightarrow\Pll}}
\newcommand{\Peenn}{\im{\Pee \kern -0.35em \rightarrow\Pnn}}
\newcommand{\Peeqq}{\im{\Pee \kern -0.35em \rightarrow\Pqq}}
\newcommand{\Peehad}{\im{\Pee \kern -0.35em \rightarrow\Phad}}
\newcommand{\Peeff}{\im{\Pee \kern -0.35em \rightarrow\Pff}}
\newcommand{\PeeTT}{\im{\Pee \kern -0.35em \rightarrow\PTT}}
\newcommand{\PeeWW}{\im{\Pee \kern -0.35em \rightarrow\PWW}}
\newcommand{\PeeZZ}{\im{\Pee \kern -0.35em \rightarrow\PZZ}}
\newcommand{\Pffff}{\im{\Pff\Pff}}
\newcommand{\Pqqqq}{\im{\Pqq\Pqq}}

\newcommand{\aqcd}{\im{\alpha_S}}

\newcommand{\GF}{\im{G_{\mathrm{F}}}}

\begin{document}

\begin{titlepage} 
\noindent                
{\Large October 1, 2002 \hfill UCD-EXPH/021001\\
 $\phantom{0}$          \hfill  hep-ex/0210003}
\begin{center}
\includegraphics[width=4cm]{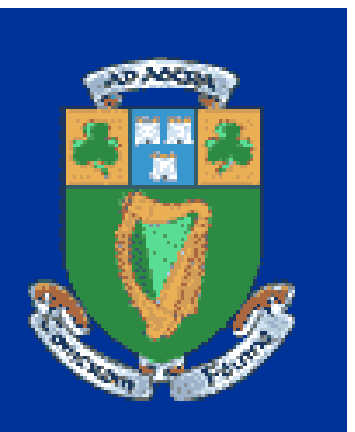}
\vskip 1cm            
{\Huge\bf Electroweak Physics\\}
\vskip 1cm
{\Large {\bf Martin W. Gr\"unewald}\\[20pt]
        University College Dublin\\
        Department of Experimental Physics\\
        Belfield, Dublin 4\\
        Ireland\\}
\vfill
{\bf Abstract} \\[10pt]
\end{center}
{\large 
  
  The status of precision electroweak measurements as of summer 2002
  is reviewed. The recent results on the anomalous magnetic moment of
  the muon and on neutrino-nucleon scattering are discussed.
  Precision results on the electroweak interaction obtained by the
  experiments at the SLC, LEP and TEVATRON colliders are presented.
  The experimental results are compared with the predictions of the
  minimal Standard Model and are used to constrain its parameters,
  including the mass of the Higgs boson.  The final LEP results on the
  direct search for the Higgs boson of the Standard Model are
  presented.

}\vfill
\begin{center}
\large\em{
Plenary talk presented at the 31st International Conference on High Energy
  Physics, \\
 Amsterdam, The Netherlands, July 24-31, 2002}
\end{center}

\end{titlepage}
\clearpage
\begin{titlepage}
$ $
\end{titlepage}
\clearpage

\setcounter{page}{1}

\title{Electroweak Physics}

\author{Martin~W.~Gr\"unewald\address{
          Department of Experimental Physics, 
          University College Dublin, 
          Belfield, Dublin 4, Ireland}%
          \thanks{E-mail: Martin.Grunewald@cern.ch}
}
       
\begin{abstract}
  
  The status of precision electroweak measurements as of summer 2002
  is reviewed. The recent results on the anomalous magnetic moment of
  the muon and on neutrino-nucleon scattering are discussed.
  Precision results on the electroweak interaction obtained by the
  experiments at the SLC, LEP and TEVATRON colliders are presented.
  The experimental results are compared with the predictions of the
  minimal Standard Model and are used to constrain its parameters,
  including the mass of the Higgs boson.  The final LEP results on the
  direct search for the Higgs boson of the Standard Model are
  presented.

\end{abstract}

\maketitle

\section{INTRODUCTION}

In this paper the current status of the precision measurements of
observables in electroweak physics are reviewed. In increasing energy
scale, the main measurements consist of: the muon anomalous magnetic
moment measured by the Brookhaven collaboration E821, neutrino-nucleon
scattering measured by the NuTeV collaboration, the precise results on
Z boson properties determined at the electron-positron colliders SLC
and LEP-1, and results on W-boson properties determined at the
TEVATRON and at LEP-2.

The measurements are compared with the expectations calculated in the
framework of the minimal Standard Model (SM).  Radiative corrections
are extracted from the measurements and tested quantitatively, in
particular by comparing predicted masses of top quark and W boson
derived from radiative corrections with the results of the direct
measurements.  Based on the complete set of measurements, constraints
are set on the mass of the SM Higgs boson.  The final results on the
direct search for the SM Higgs boson at LEP-2 are also presented.

\section{ANOMALOUS MAGNETIC MOMENT OF THE MUON}

\begin{figure}[tbp]
\begin{center}
  \includegraphics[clip=true,bb=0 0 245
  170,width=0.8\linewidth]{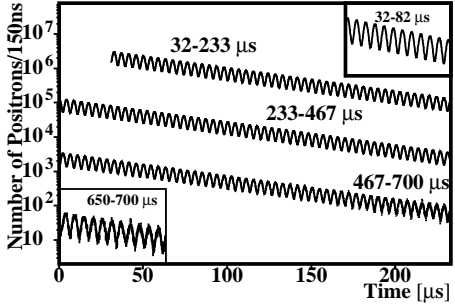} \vskip -1cm
\caption{Counting rate recorded by E821 as a function of the muon
decay time, modulated by the spin precession frequency $\omega$. }
\label{fig:g-2-wiggles}
\end{center}
\end{figure}

The Brookhaven experiment E821 measures the anomalous magnetic moment
of the muon, $a_\mu$, by measuring the muon spin precession frequency,
$\omega$, in an external magnetic field, B:
\begin{eqnarray}
(g_\mu -2)/2 & = & a_\mu ~ = ~ (\omega m_\mu c)/(eB)\,,
\end{eqnarray}
where the magnetic field B is measured with a proton NMR probe.  Based
on the data collected in 1999~\cite{BNL-E821-1999}, consisting of
nearly $10^9$ positrons from $\mu^+$ decay, the time spectrum shown in
Figure~\ref{fig:g-2-wiggles} is obtained, yielding the result:
\begin{eqnarray}
10^{10}a_\mu & = & 11\,659\,202 \pm 15\,,
\end{eqnarray}
to be compared with the expected value of $11\,659\,160\pm7$ in units
of $10^{-10}$ calculated in the framework of the minimal SM.  The
expectation being lower by about 2.6 standard deviations compared to
the experimental result caused great excitement because shifts in
$a_\mu$ could signal new physics~\cite{OLD+NEW}, \eg, supersymmetry.

\begin{figure}[tbp]
\begin{center}
\includegraphics[width=0.23\linewidth]{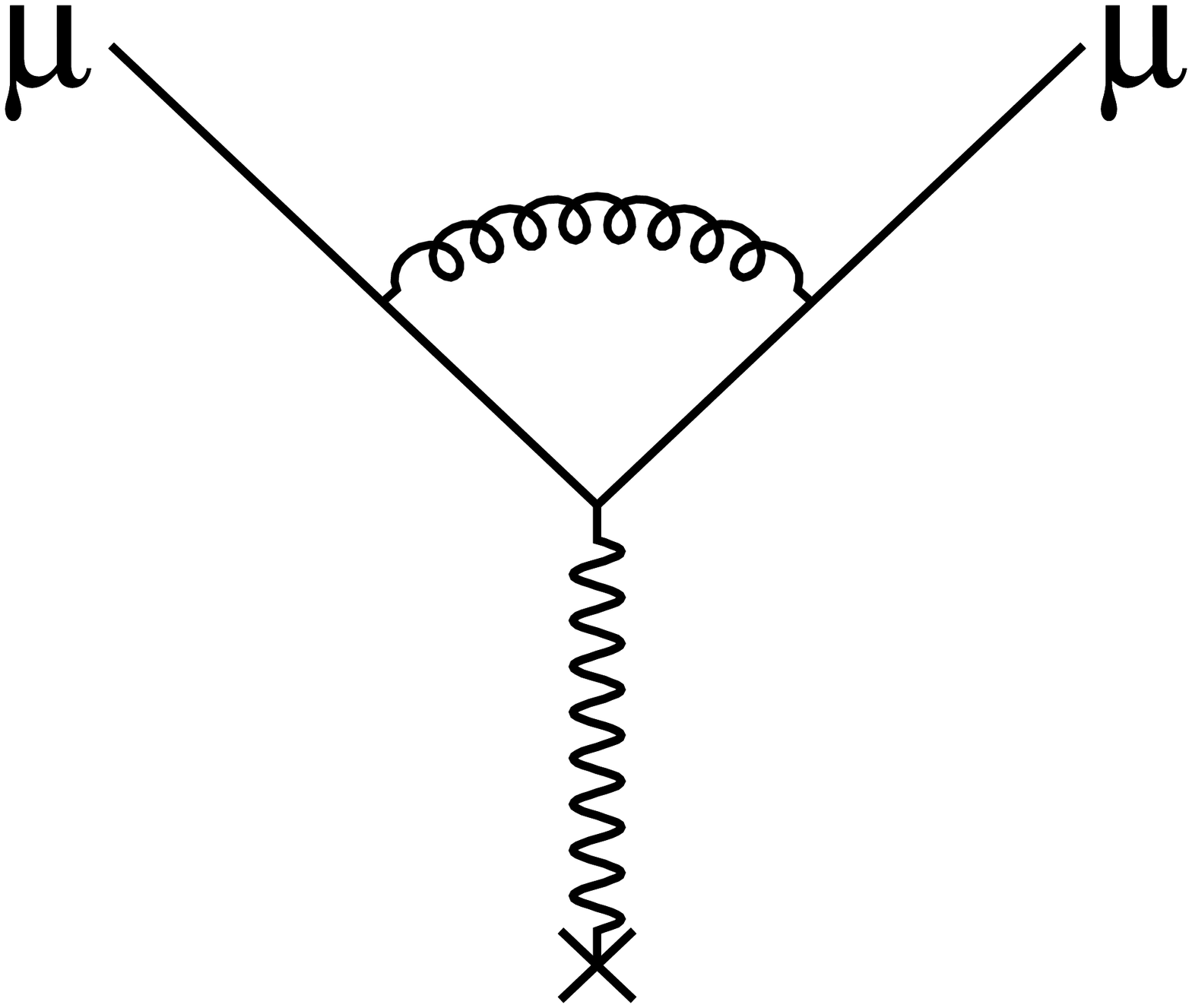}
\includegraphics[width=0.23\linewidth]{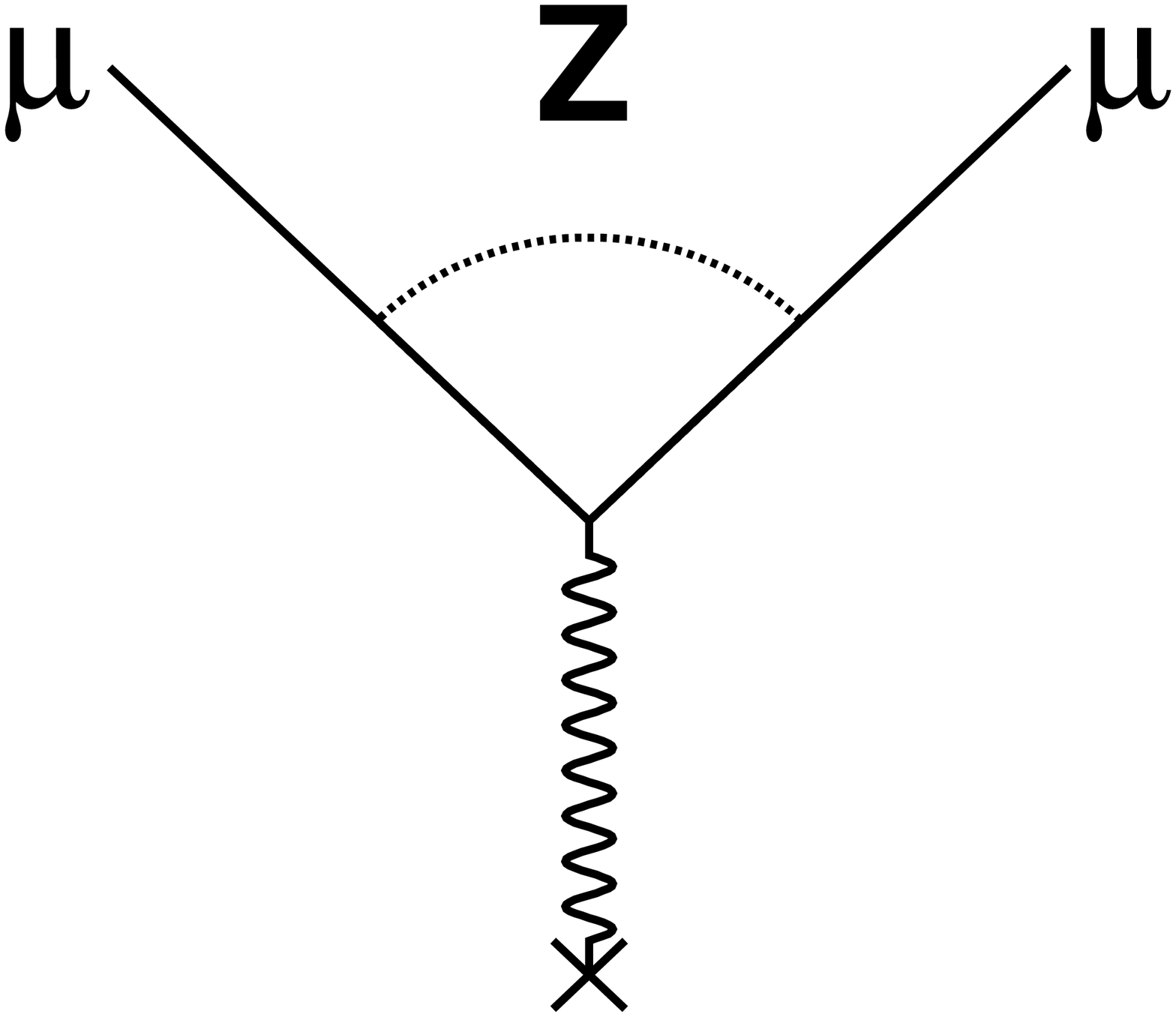}
\includegraphics[width=0.23\linewidth]{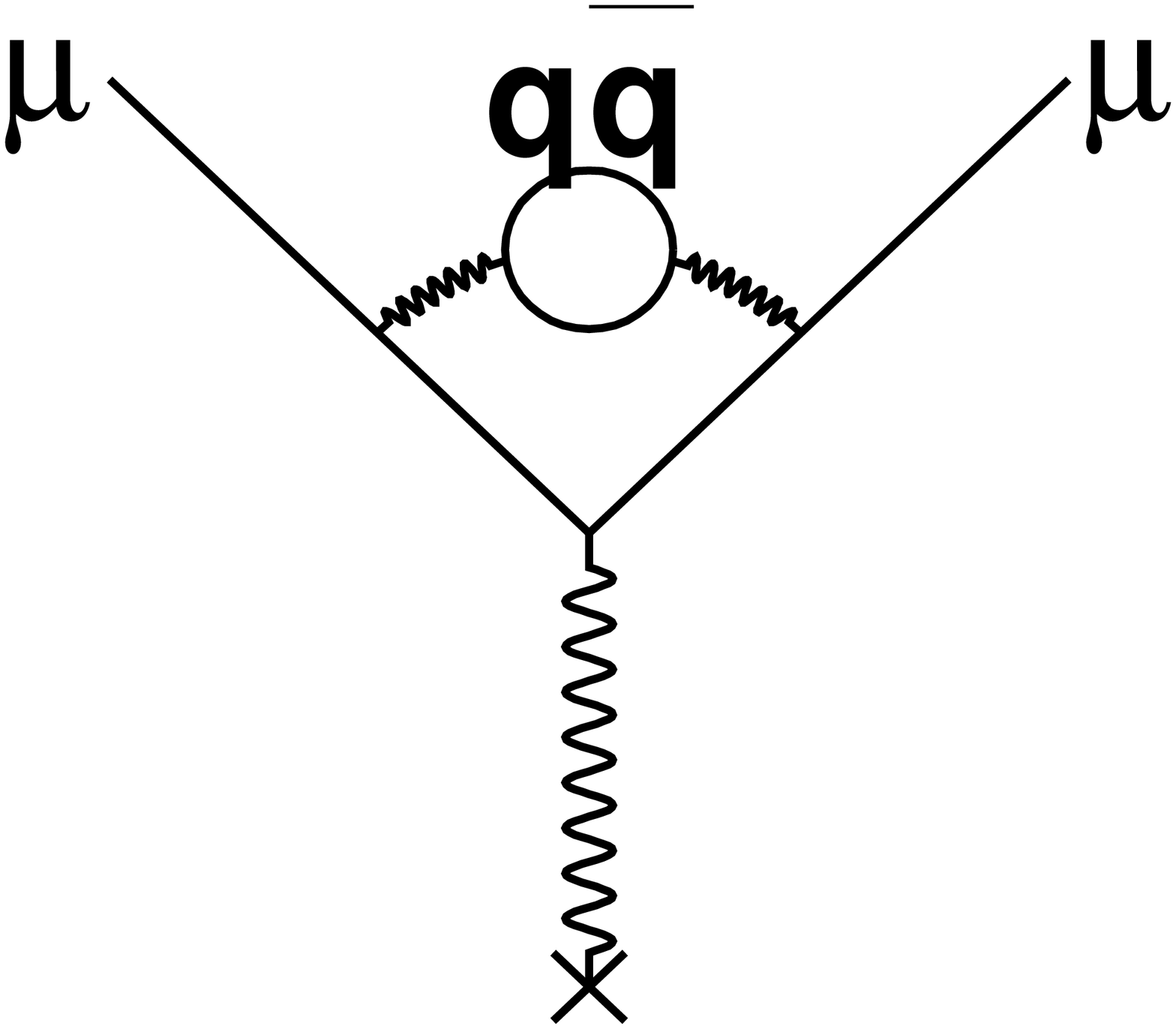}
\includegraphics[width=0.23\linewidth]{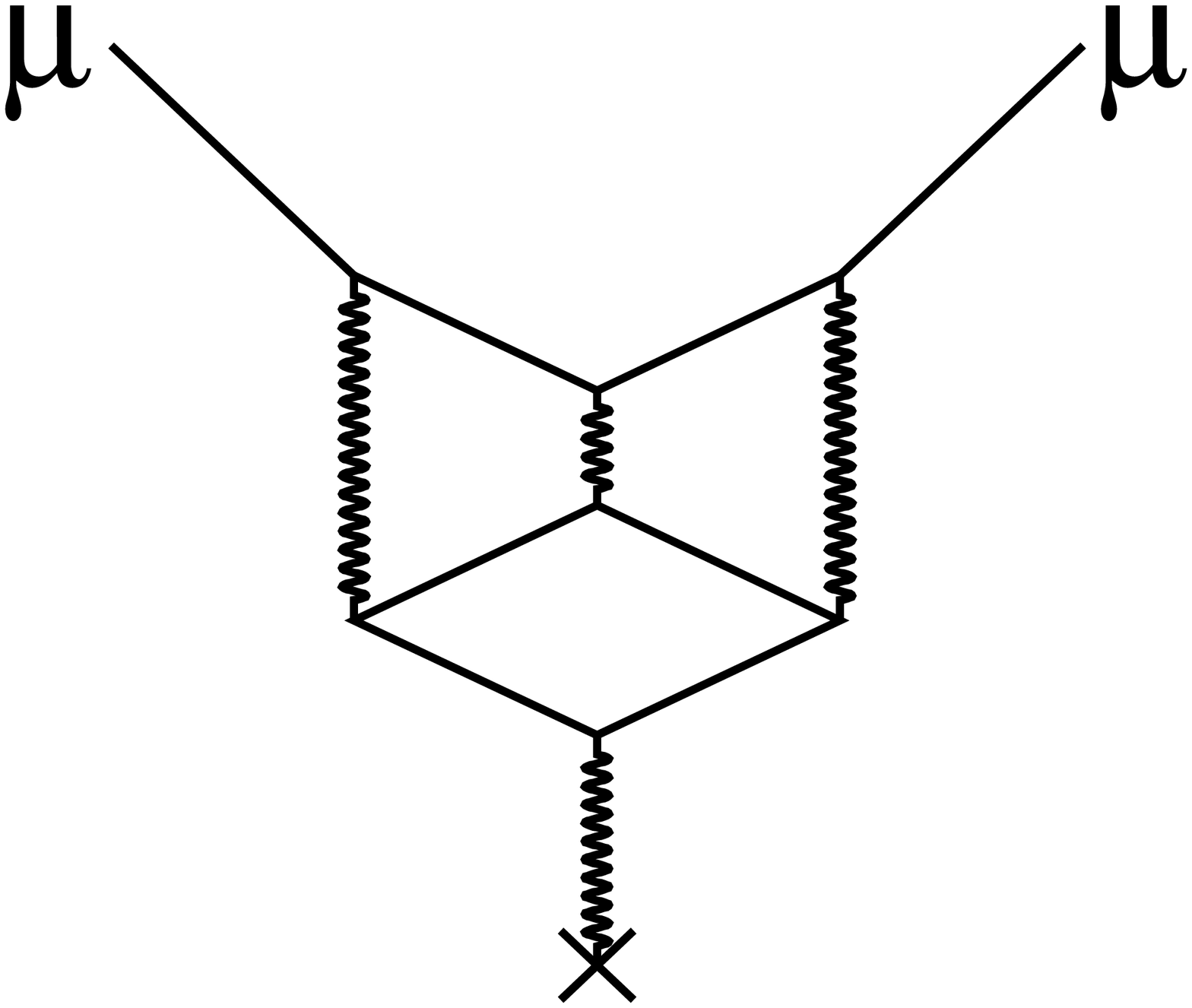}
\vskip -1cm
\caption{Feynman diagrams contributing to the anomalous magnetic
moment of the muon. From left to right: QED, weak, hadronic vacuum
polarisation and the light-by-light contribution.}
\label{fig:g-2-SM}
\end{center}
\end{figure}

The SM expectation contains four contributions, graphically shown in
Figure~\ref{fig:g-2-SM}.  The QED part is calculated in a perturbation
series in terms of powers of $\alpha/\pi$, the first order term being
the Schwinger term $\alpha/2\pi$. This part, +11658471, is by far the
dominant part, but still calculated with negligible uncertainty.
Likewise, the weak part, +15 units, is known with negligible
uncertainty, while it's size is comparable to the experimental
uncertainty.  The uncertainty on the prediction of $a_\mu$ is fully
dominated by the hadronic vacuum polarisation, where estimates for the
first-order contribution range from 692 to 699 units, with
uncertainties of about 7 units.  The E821 collaboration has chosen the
calculation yielding the result at the lower edge of this range.  The
second-order contribution is small, $-10\pm1$ units.  The hadronic
light-by-light scattering contribution had been calculated to be
$-8\pm4$ units.  Here, a sign error in the dominant contribution, the
pion pole, has been discovered~\cite{SIGN}.  When corrected, this
contribution changes to $+8\pm2$ units; the full SM prediction
increases to $11\,659\,177\pm7$ units, reducing the difference between
the experimental measurement and the theoretical calculation to 1.6
standard deviations.

On the last day of this conference, a new result of the E821
collaboration was presented based on the data collected in the year
2000~\cite{g-2-EXP,BNL-E821-2000}, corresponding to an additional
$4\cdot10^9$ $\mu^+$ decays:
\begin{eqnarray}
10^{10}a_\mu & = & 11\,659\,204 \pm 8,
\end{eqnarray}
which dominates the new world average value of:
\begin{eqnarray}
10^{10}a_\mu & = & 11\,659\,203 \pm 8.
\end{eqnarray}
The difference between theory and experiment is again at the level of
2.6 standard deviations.

Because of its large uncertainty, the first-order hadronic vacuum
polarisation is under continued scrutiny.  At this conference, new
calculations were presented~\cite{g-2-TH}, predicting contributions
even lower than the value of 692 used by E821.  Clearly more work is
needed to understand the large spread in the calculations, which
should agree much better as they are all based on the same low energy
data for electron-positron annihilations and $\tau$ decays into
hadrons~\cite{Hagiwara,DavierHocker}.

\section{NEUTRINO NUCLEON SCATTERING}

\begin{figure}[tb]
\begin{center}
\includegraphics[width=0.3\linewidth]{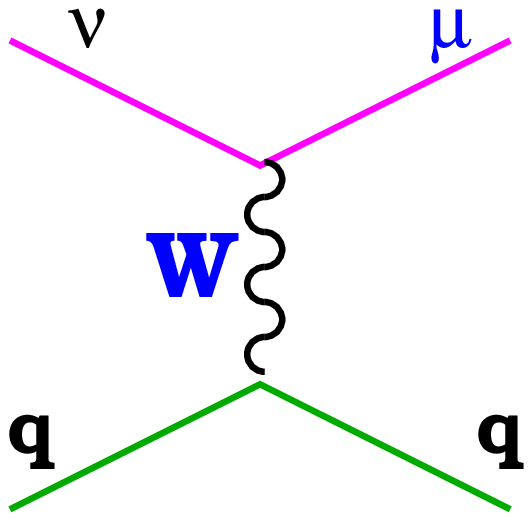}
\hfill
\includegraphics[width=0.3\linewidth]{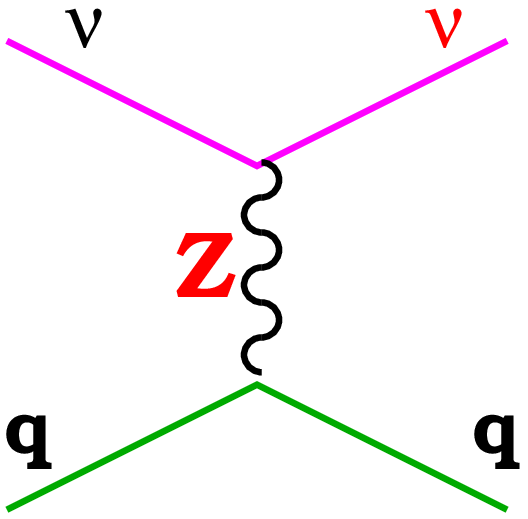}
\vskip -1cm
\caption{Feynman diagrams in $\nu$N scattering at parton
  level. Left: CC, right: NC interactions.}
\label{fig:nutev-t}
\end{center}
\end{figure}

The NuTeV collaboration measures the electroweak mixing angle in
$t$-channel neutrino-nucleon scattering as shown in
Figure~\ref{fig:nutev-t}, involving charged current (CC) and neutral
current (NC) reactions.  Using for the first time both a neutrino and
an anti-neutrino beam with high statistics, it is possible to exploit
the Paschos-Wolfenstein relation~\cite{PWR}:
\begin{eqnarray}
R_- & = & \frac{\sigma_{NC}(\nu)-\sigma_{NC}(\bar\nu)}
               {\sigma_{CC}(\nu)-\sigma_{CC}(\bar\nu)} \\
& = & 4 g^2_{L\nu}\sum_{u,d}\left[g^2_{Lq} - g^2_{Rq}\right] \\
& = & \rho_\nu\rho_{ud}\left[1/2 -
               \sin^2\theta_W^{on-shell}\right]\,,
\end{eqnarray}
where the sum runs over the valence quarks u and d. This relation
holds for iso-scalar targets and up to small electroweak radiative
corrections. Thus $R_-$ is a measurement of the on-shell electroweak
mixing angle.  In the ideal case this measurement is insensitive to
the effects of sea quarks, which cancel. Charm production, uncertain
due to charm mass effects, enter only through CC scattering off
valence d quarks, a CKM suppressed process.

\begin{figure}[tbp]
\begin{center}
\includegraphics[clip=true,bb=500 60 645 600,angle=-90,width=0.9\linewidth]{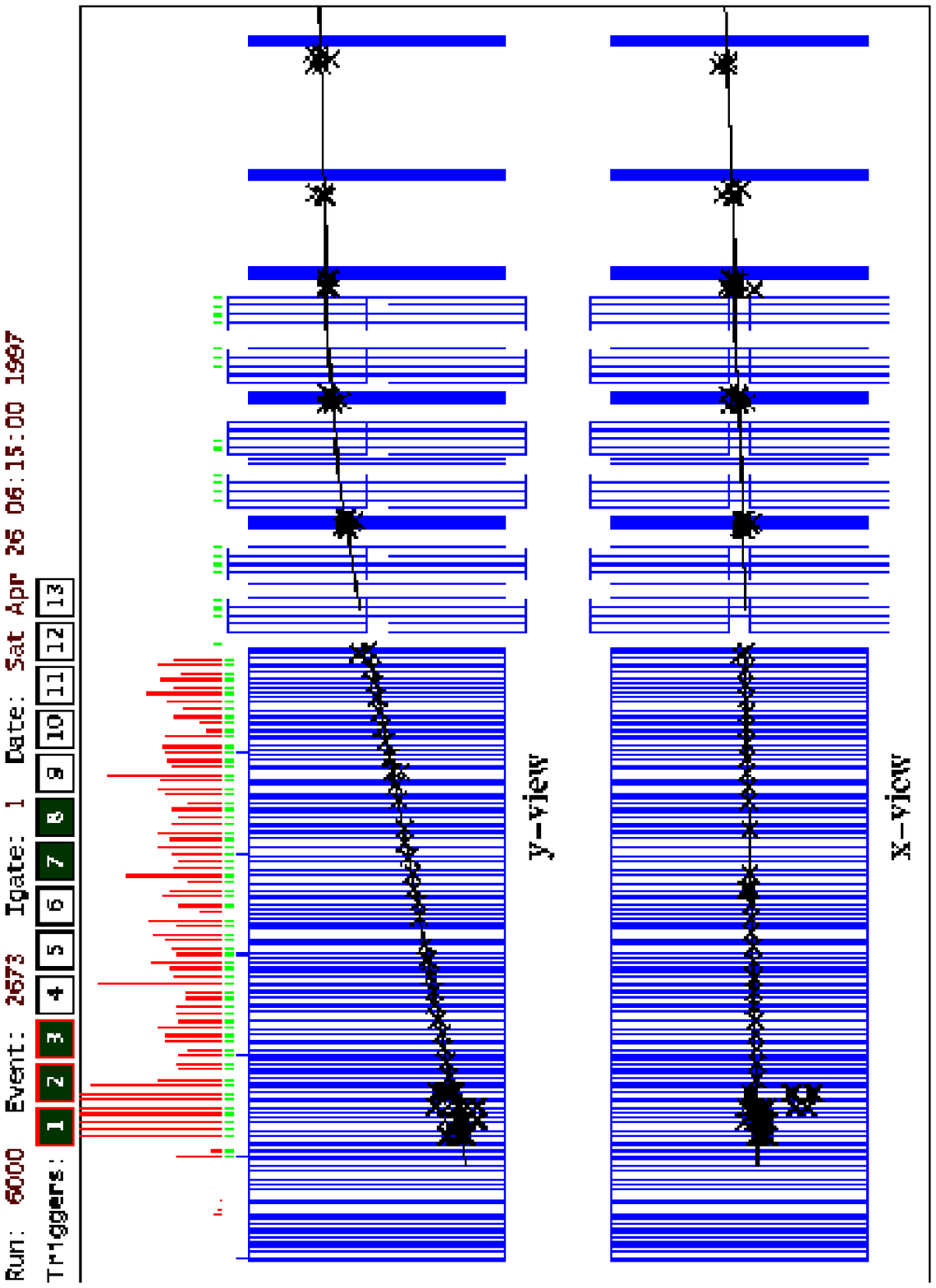}
\includegraphics[clip=true,bb=500 60 645 600,angle=-90,width=0.9\linewidth]{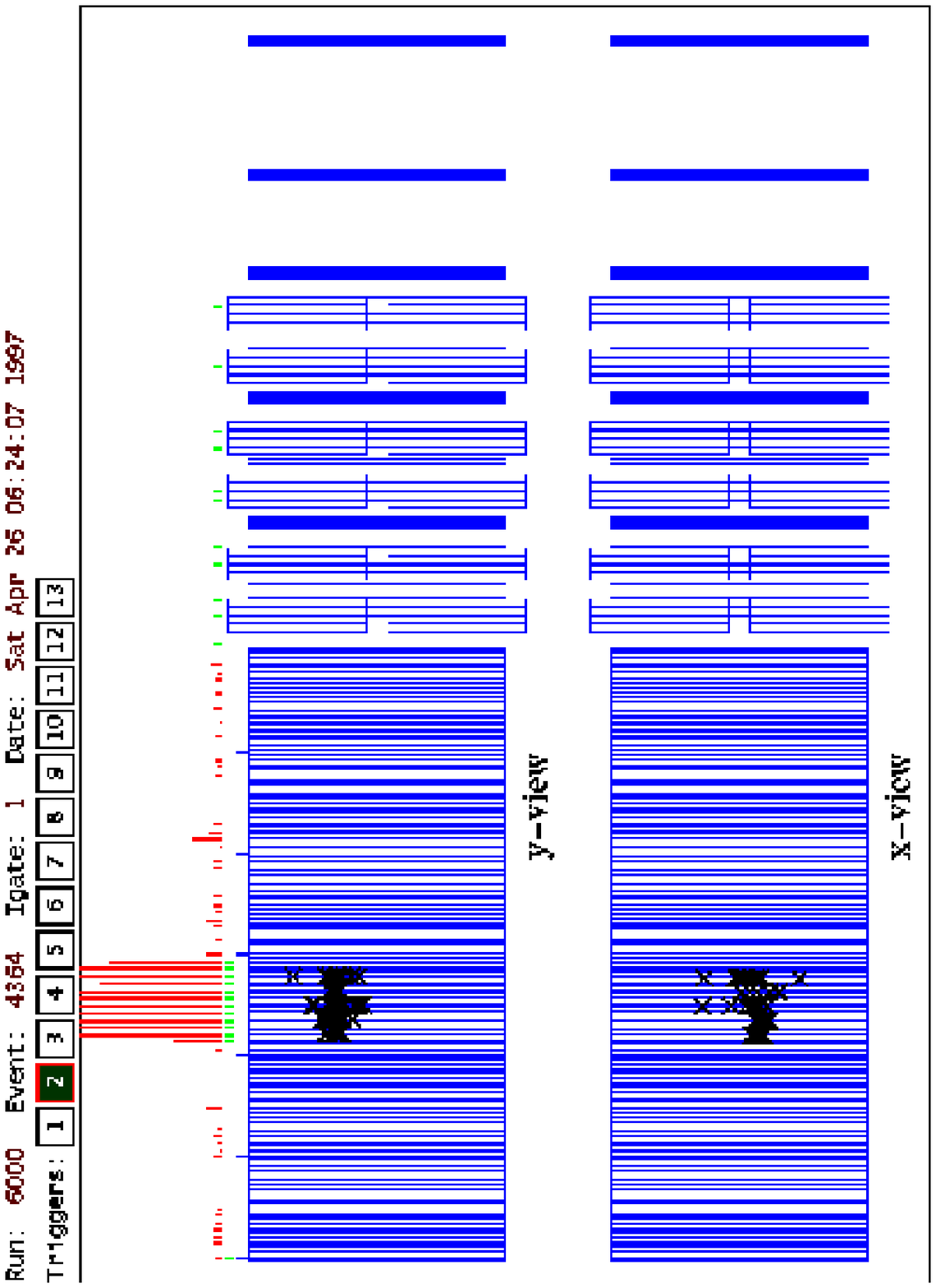}
\vskip -0.5cm
\caption{Events observed in the NuTeV detector. Top: CC
  event, bottom: NC event.}
\label{fig:nutev-events}
\end{center}
\end{figure}

Using a muon (anti-) neutrino beam, CC reactions contain a primary
muon in the final state, while NC reactions do not. As shown in
Figure~\ref{fig:nutev-events}, the muon as a minimum ionising particle
traverses the complete detector, while the hadronic shower is confined
in a small target volume. The length of the event thus discriminates
clearly between CC and NC events.


\begin{figure}[tbp]
\begin{center}
\vskip -0.75cm
\includegraphics[width=0.9\linewidth]{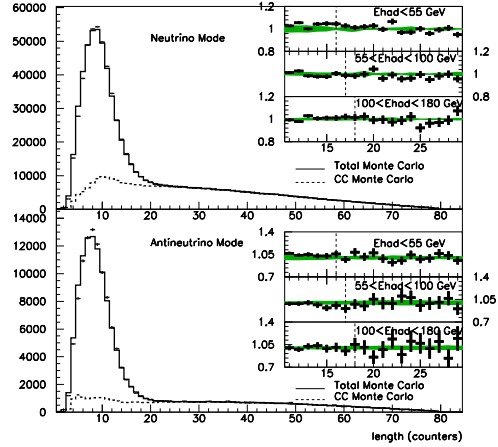}
\vskip -1cm
\caption{Distribution of event lengths.}
\label{fig:nutev-length}
\end{center}
\end{figure}

The distributions of event lengths as observed for neutrino and
anti-neutrino beams are shown in Figure~\ref{fig:nutev-length}. In
total, close to 2 million events were recorded by the NuTeV
collaboration, 1167K CC and 457K NC events with neutrino beams, and
250K CC and 101K NC events with anti-neutrino beams. The separation
between CC and NC events is dependent on the energy of the hadronic
shower and ranges from 16 to 18 in units of counters (equivalent to
10~cm of steel) as indicated in the inserts of
Figure~\ref{fig:nutev-length}.

\begin{figure}[tbp]
\begin{center}
\includegraphics[width=0.8\linewidth]{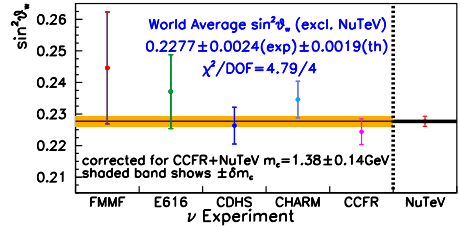}
\vskip -1cm
\caption{Comparison of results on the on-shell electroweak mixing
  angle in $\nu$N experiments.}
\label{fig:nutev-world}
\end{center}
\end{figure}

In order to extract $R_-$ from the measured distributions, a Monte
Carlo simulation of the spectra of the \hbox{(anti-)} neutrino beams,
radiative corrections and detector response is used. In terms of the
on-shell electroweak mixing angle, NuTeV's final results
reads~\cite{NuTeV}:
\begin{eqnarray}
\sin^2\theta_W^{on-shell} & \equiv & 1-\MW^2/\MZ^2 \\
& = & 0.2277 \pm 0.0013  \pm 0.0009  \nonumber \\
&   & - 0.00022 \frac{\MT^2-(175~\GeV)^2}{(50~\GeV)^2} \nonumber \\
&   & + 0.00032 \ln(\MH/150~\GeV)\,, \nonumber
\end{eqnarray}
where the first error is statistical and the second is systematic.
Here $\rho=\rho_{SM}$ is assumed. This result is in very good
agreement with the previous world average of
$0.2277\pm0.0024(exp.)\pm0.0019(theo)$ updated for the latest charm
mass~\cite{nNWorld}, as shown in Figure~\ref{fig:nutev-world}.

The two contributions to the total systematic uncertainty of 0.0009
are about equal, 0.0006 each for experimental systematics and for
modelling. The experimental systematics are dominated by the
uncertainty on the (anti-)electron neutrino flux, as for such beams
both CC and NC interactions lead to final states without primary
muons.  The model systematics are dominated by charm production and
the strange-quark sea, effects which are much reduced compared to
previous single-beam experiments.  With a statistical error of 0.0013
and a total systematic error of 0.0009, NuTeV's final result is
statistics limited.

\begin{figure}[tbp]
\begin{center}
\includegraphics[width=0.8\linewidth]{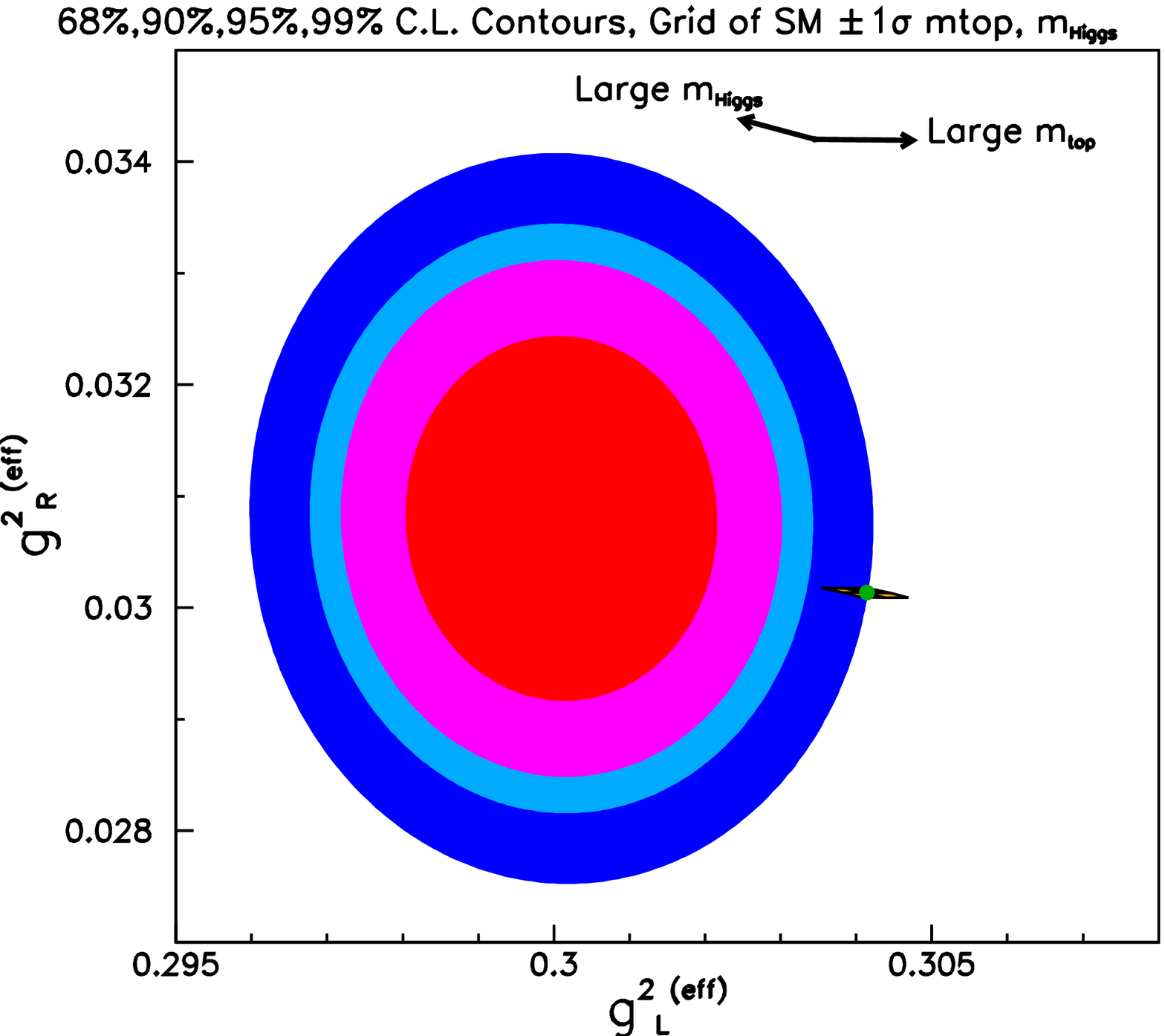}
\includegraphics[width=0.8\linewidth]{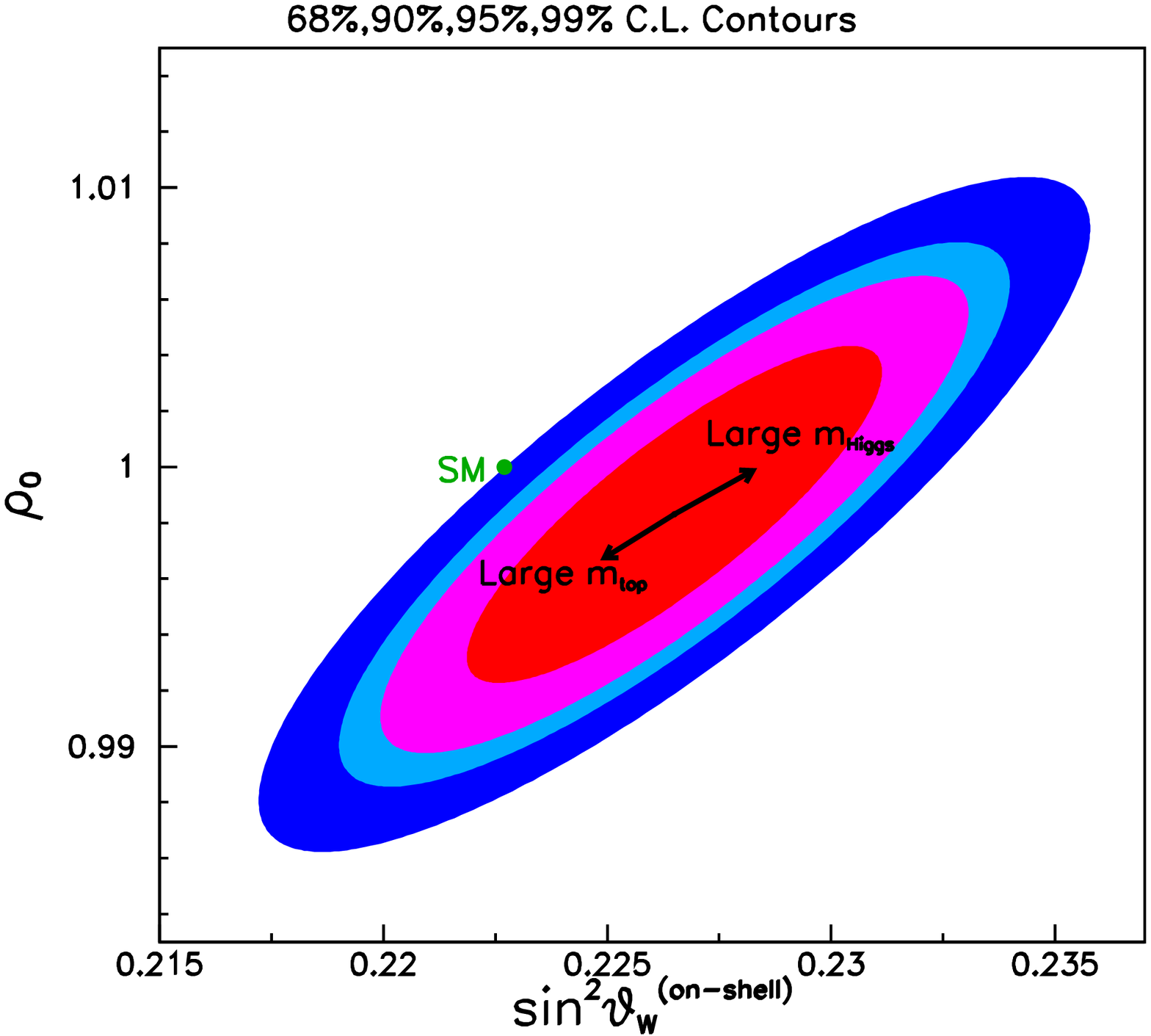}
\vskip -1cm
\caption{NuTeV result in the plane of effective right- and left-handed
  couplings (top) and of on-shell angle versus $\rho$-scale factor
  $\rho_0$. }
\label{fig:nutev-2d}
\end{center}
\end{figure}

Also this result caused a great deal of excitement, as the global SM
analysis of all electroweak measurements, presented later in this
paper, predicts a value of $0.2227\pm0.0004$ for the on-shell
electroweak mixing angle, showing a deviation from the NuTeV result at
the level of 3.0 standard deviations. 

In a more model-independent analysis, the NuTeV result is also
interpreted in terms of effective left- and right-handed couplings,
shown in Figure~\ref{fig:nutev-2d}, defined as: $g^2_X(eff) =
4g^2_{L\nu}\sum_q g^2_{Xq} $ for $X=L,R$.  Here the deviation is
confined to the effective left-handed coupling product.  Modifying all
$\rho$ parameters by a scale factor $\rho_0$, also shown in
Figure~\ref{fig:nutev-2d}, shows that either $\rho_0$ or the mixing
angle, but not both, could be in agreement with the SM.  Assuming the
electroweak mixing angle to have it's expected value, the change in
the $\rho$ factors can be absorbed in $\rho_\nu$, \ie, interpreted as
a change in the coupling strength of neutrinos, then lower than
expected by about $(1.2\pm0.4)\%$.  A similar trend is observed with
the neutrino coupling as measured by the invisible width of the Z
boson at LEP-1, yielding a much less significant deficit of
$(0.5\pm0.3)\%$ in $\rho_\nu$.

To date, various explanations ranging from old and new physics effects
have been put forward and reviewed at this
conference~\cite{PrecEW-TH}.  Some old physics effects are:
theoretical uncertainties in PDFs, iso-spin violating PDFs,
quark-antiquark asymmetries for sea quarks, nuclear shadowing
asymmetries between W and Z interactions, etc. Some new physics
effects are: a new heavy Z boson, contact interactions, lepto-quarks,
new fermions, neutrinos oscillations, etc. Most of the old and new
physics effects are, however, severely constrained by NuTeV itself or
other precision electroweak measurements, thus cannot explain the full
effect.  As was pointed out at this conference, PDFs should be
investigated, in particular the partly leading-order analysis employed
by NuTeV should be assessed and eventually improved to next-to-leading
order.

\section{Z-BOSON PHYSICS}

\begin{figure}[tbp]
\begin{center}
\includegraphics[width=0.8\linewidth]{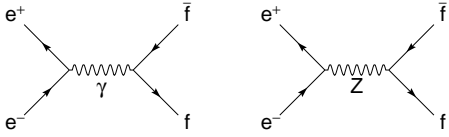}
\vskip -1cm
\caption{Feynman diagrams in $\Pff$ production. Left: photon
  exchange; right: Z boson exchange.}
\label{fig:eesff}
\end{center}
\end{figure}

\begin{figure}[tbp]
\begin{center}
\includegraphics[width=0.9\linewidth]{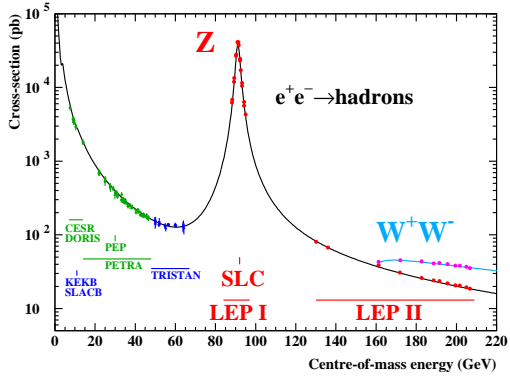}
\vskip -1cm
\caption{Hadron production in $\Pee$ annihilation.}
\label{fig:ee-had}
\end{center}
\end{figure}

In the last decade, electron-positron annihilations at high energies
have allowed to measure precisely a wealth of electroweak observables
related to on-shell Z bosons decaying to fermion-antifermion pairs.
The lowest-order Feynman diagrams contributing to fermion-pair
production are shown in Figure~\ref{fig:eesff}.
Figure~\ref{fig:ee-had} shows the total cross section for hadron
production as a function of the $\Pee$ centre-of-mass energy.
Prominent features are the 1/s fall off at low energies due to virtual
photon exchange, and the dominant Z resonance at $91~\GeV$.  Also
shown is the threshold for $\PWW$ production around $160~\GeV$.

The properties of the Z boson are measured precisely by the SLC
experiment SLD, and the LEP experiments ALEPH, DELPHI, L3 and OPAL.
For example~\cite{LEPLS}, the mass and width of the Z boson are
$\MZ=91187.5\pm2.1~\MeV$, and $\GZ=2495.2\pm2.3~\MeV$.  The number of
light neutrinos species is found to be $2.9841\pm0.0083$, about 1.9
standard deviations smaller than three as already observed in the
previous section.


\subsection{Photon-Z Interference}

\begin{figure}[tbp]
\begin{center}
\includegraphics[width=0.8\linewidth]{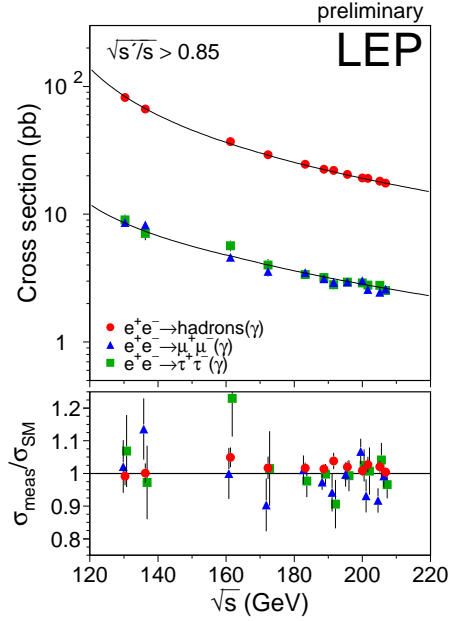}
\vskip -1cm
\caption{Measured cross sections for $\Pff$ production at energies above
  the Z-pole, compared to the SM expectation. }
\label{fig:lep2-xsec}
\end{center}
\end{figure}

\begin{figure}[tbp]
\begin{center}
\includegraphics[width=0.8\linewidth]{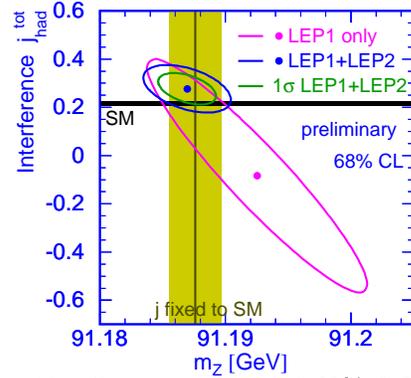}
\vskip -1.5cm
\caption{Contour curves of 68\% C.L. in the plane of Z mass versus
interference term for the hadronic final state. 
}
\label{fig:smat}
\end{center}
\end{figure}

Owing to the increase in centre-of-mass energy at LEP-2, fermion-pair
production is now tested at energies much higher than the Z pole.  The
results from the LEP experiments, still preliminary, are combined
taking common systematic uncertainties into account~\cite{LEP2-2F}.
The combined cross sections are shown in Figure~\ref{fig:lep2-xsec}
and compared to the SM expectation.  Besides searches for new physics,
these data are used to determine the interference term between photon
and Z-boson exchange for the hadronic final state. This is interesting
and important because this interference term is fixed to its SM value
when extracting the Z-boson mass quoted above from the LEP-1 Z-pole
measurements. The situation is shown graphically in
Figure~\ref{fig:smat}. Relaxing the SM constraint on the hadronic
interference term, the LEP-1 data alone shows a large correlation with
$\MZ$, increasing the uncertainty on $\MZ$. Including the LEP-2 data
constrains the interference term. The error on $\MZ$ is reduced close
to that obtained in the LEP-1 analysis with fixed interference term.
Similar analysis are possible with data collected below the Z pole,
\eg, by TOPAZ~\cite{TOPAZ} and VENUS~\cite{VENUS} at TRISTAN, albeit
with larger uncertainties.


\begin{figure}[tbp]
\begin{center}
\includegraphics[width=0.8\linewidth]{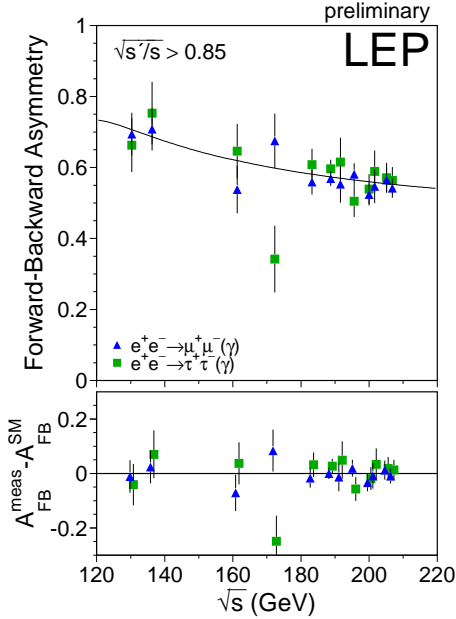}
\vskip -1cm
\caption{Measured forward-backward asymmetries for $\Pff$ at
  energies above the Z-pole, compared to the SM expectation.}
\label{fig:lep2-afb}
\end{center}
\end{figure}

The forward-backward asymmetries as measured at energies above the
Z-pole are shown in Figure~\ref{fig:lep2-afb}. Analyses are currently
ongoing to extract the $\gamma$/Z interference terms for leptons and
for forward-backward asymmetries.

\subsection{Leptonic Polarisation Asymmetries}

\begin{figure}[tbp]
\begin{center}
\includegraphics[width=0.8\linewidth]{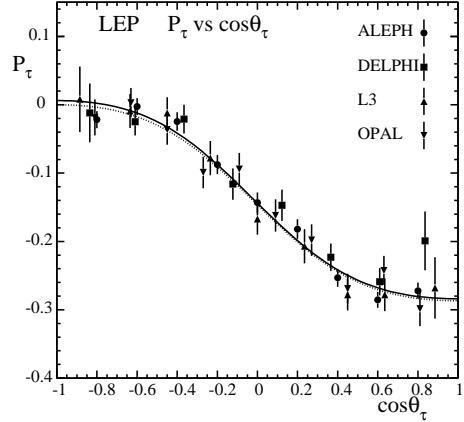}
\vskip -1cm
\caption{Tau polarisation as a function of the polar scattering angle
  in tau-pair production at LEP-1. The results of a fit to the data,
  with or without assuming e-$\tau$ universality, are shown as the
  dashed and solid line. }
\label{fig:lep1-ptau}
\end{center}
\end{figure}

In terms of the effective vector and axial-vector coupling constants,
$\gvf$ and $\gaf$, the asymmetry parameter $\Af$ is defined as:
\begin{eqnarray}
\Af & = & 2 \frac{\gvf/\gaf}{1+(\gvf/\gaf)^2} \,.
\end{eqnarray}
The leptonic asymmetry parameter is measured at SLC and at LEP-1 in
various processes.  Assuming lepton universality, well supported by
the experimental results, the following final results are obtained
when combining the experiments:
\begin{eqnarray}
\Al & = & 0.1512\pm0.0042 \quad \hbox{f/b asymmetry}\\
\Al & = & 0.1465\pm0.0033 \quad \hbox{$\tau$ polarisation}\\
\Al & = & 0.1513\pm0.0021 \quad \hbox{l/r asymmetries}\,,
\end{eqnarray}
in good agreement and with a combined value of:
\begin{eqnarray}
\Al & = & 0.1501\pm0.0016 \,.
\end{eqnarray}
As an example, the measurement of of the $\tau$ polarisation as a
function of polar scattering angle is shown in
Figure~\ref{fig:lep1-ptau}.  For backward scattering, zero average
helicity is expected and observed, while for forward scattering the
largest polarisation is obtained.

\subsection{Effective Coupling Constants}

The effective coupling constants of the neutral weak current are
defined as:
\begin{eqnarray}
\gvf & = & \sqrt{\rho_f}\left(T^f_3 - 2 Q_f\sin^2\theta^f_{eff}\right) \\
\gaf & = & \sqrt{\rho_f}\,    T^f_3,
\end{eqnarray}
where $T^f_3$, $Q_f$ and $\sin^2\theta^f_{eff}$ are the third
component of the weak iso-spin, the electric charge and the effective
electroweak mixing angle of fermion $f$.  While the asymmetry
parameter determines $\gvf/\gaf$, the partial Z decay width constrains
$\gvfsq+\gafsq$, allowing to disentangle $\gvf$ and $\gaf$.

\begin{figure}[tbp]
\begin{center}
\vskip -0.75cm
\includegraphics[width=0.9\linewidth]{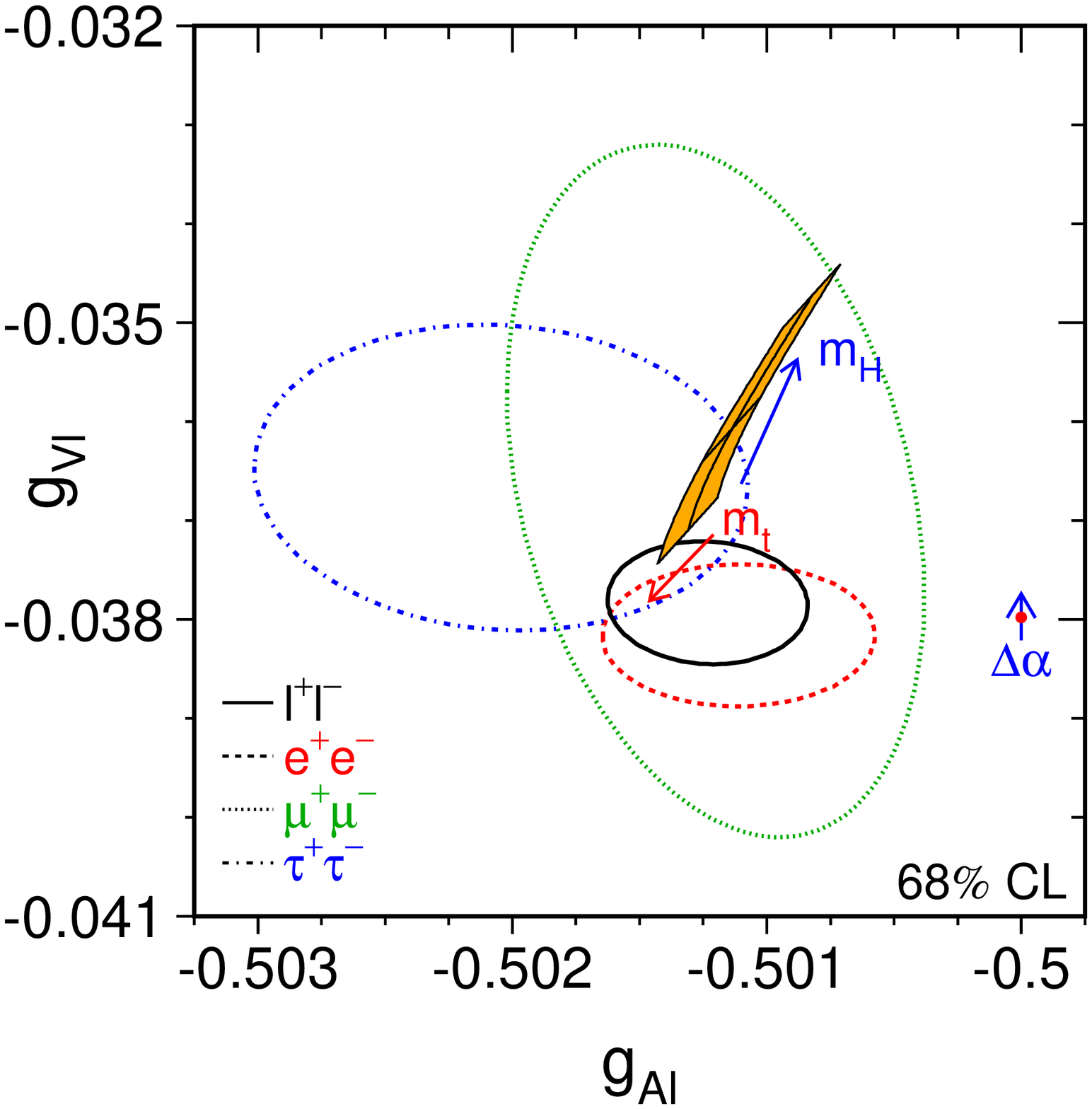}
\vskip -1.5cm
\caption{Contour curves of 68\% C.L. in the $(\gvl,\gal)$ plane. The
  SM expectation is shown as the shaded area for
  $\MT=174.3\pm5.1~\GeV$ and $\MH=300^{+700}_{-186}~\GeV$. }
\label{fig:gvga}
\end{center}
\end{figure}

The results are displayed in Figure~\ref{fig:gvga}. The three contours
for the three charged lepton species e, $\mu$ and $\tau$ show lepton
universality for both effective couplings.  Under this assumption, the
final Z-pole results are:
\begin{eqnarray}
\gvl & = & -0.03783 \pm 0.00041 \\
\gal & = & -0.50123 \pm 0.00026 \,,
\end{eqnarray}
with a correlation of $-5.9\%$.

Comparing the experimental contours with the predictions, non-QED
electroweak radiative corrections are observed with a significance of
5 standard deviations.  The results are in agreement with the
expectations calculated in the framework of the minimal SM, and show
preference for a light Higgs boson.

\subsection{Heavy Flavour Results at the Z Pole}

\begin{figure}[tbp]
\begin{center}
\vskip -0.75cm
\includegraphics[width=0.9\linewidth]{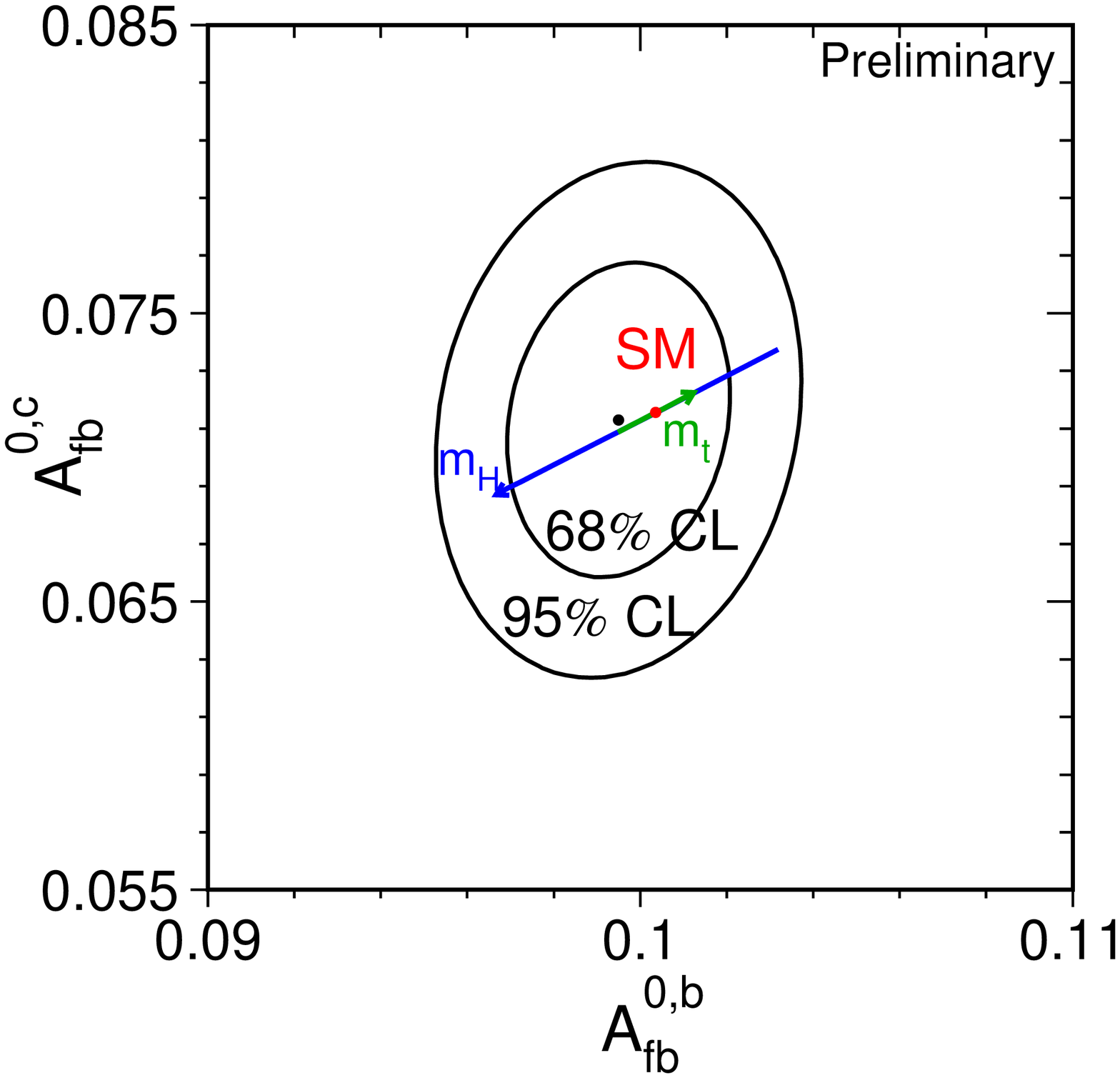}
\vskip -1.5cm
\caption{Contour curves of 68\% C.L. in the $(\Afbzb,\Afbzc)$ plane.
  The SM expectation is shown as the arrows for $\MT=174.3\pm5.1~\GeV$
  and $\MH=300^{+700}_{-186}~\GeV$.}
\label{fig:afb0-b-c}
\end{center}
\end{figure}

While the results on b- and c-quark production rates
($\Rq=\GZq/\GZhad$) are final, several measurements of heavy-flavour
asymmetries by SLD and at LEP are still preliminary, and thus is the
joint combination of all heavy flavour results. The combined
preliminary results on partial width ratios, forward-backward pole
asymmetries $\Afbzf=3/4\Ae\Af$, and forward-backward left-right
asymmetries $\Af$ are~\cite{LEP1-HF}:
\begin{eqnarray}
\Rb    & = & 0.21644\pm 0.00065\\
\Rc    & = & 0.1718 \pm 0.0031 \\
\Afbzb & = & 0.0995 \pm 0.0017 \\
\Afbzc & = & 0.0713 \pm 0.0036 \\
\Ab    & = & 0.922  \pm 0.020  \\
\Ac    & = & 0.670  \pm 0.026  \,,
\end{eqnarray}
with the largest correlation, $+0.15$, occurring between the b and c
forward-backward asymmetries as shown in Figure~\ref{fig:afb0-b-c}.
The value of the combined forward-backward b asymmetry prefers an
intermediate Higgs-boson mass of a few hundred $\GeV$.

\begin{figure}[tbp]
\begin{center}
\includegraphics[clip=true,bb=35 165 470 500,width=0.9\linewidth]{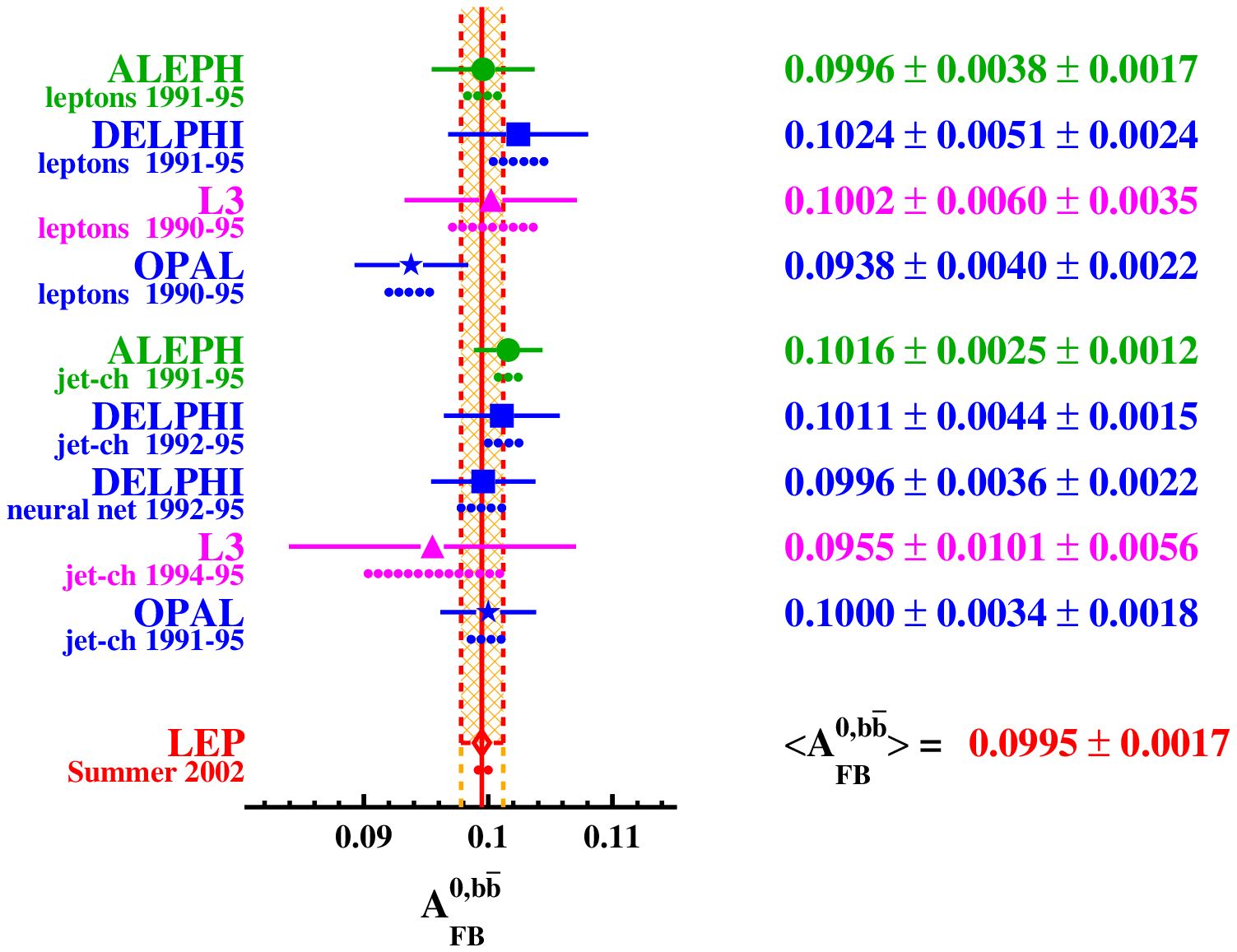}
\vskip -1cm
\caption{Measurements of $\Afbzb$ at the Z pole.}
\label{fig:hf-bar-b}
\end{center}
\end{figure}

\begin{figure}[tbp]
\begin{center}
\includegraphics[clip=true,bb=35 165 470 500,width=0.9\linewidth]{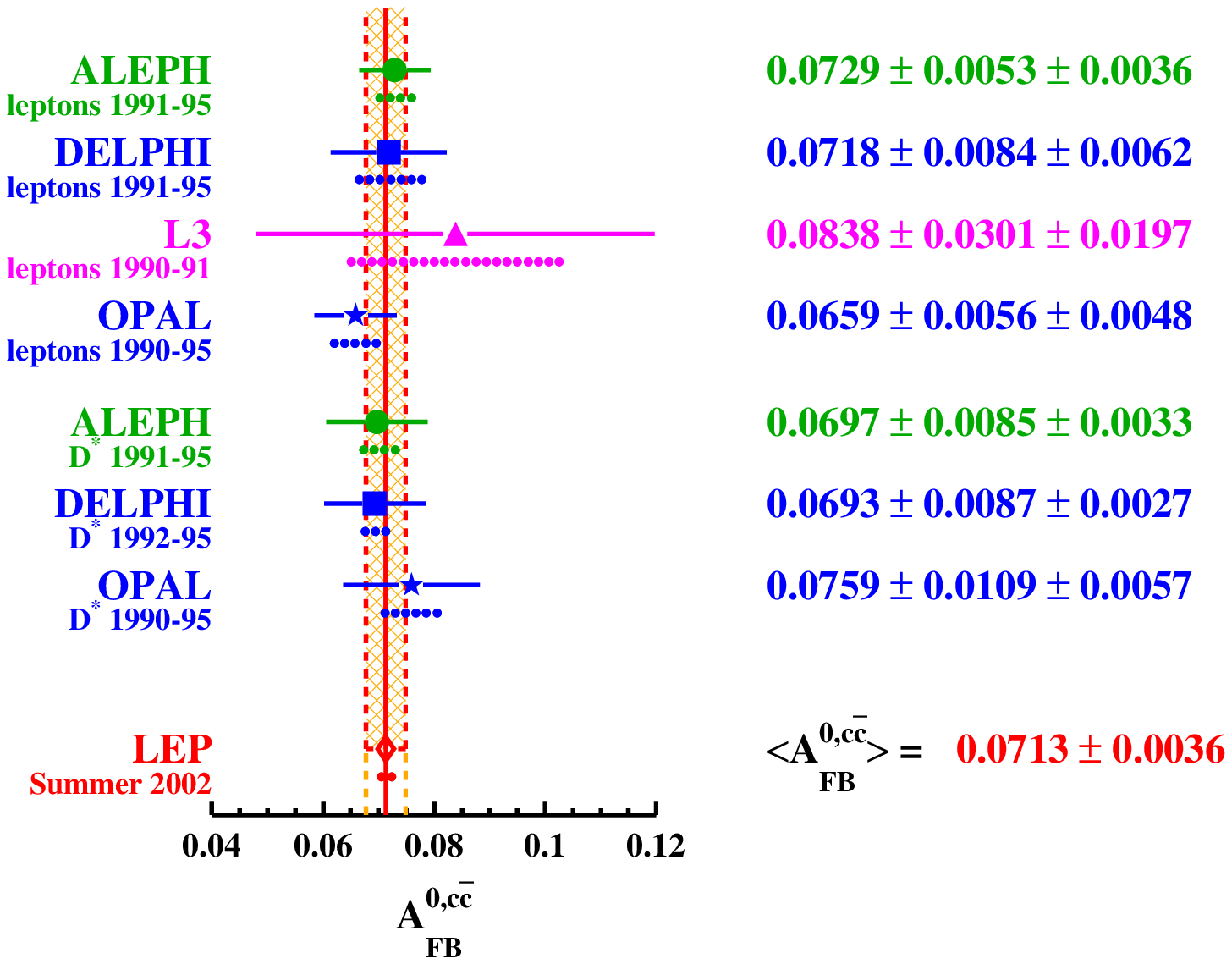}
\vskip -1cm
\caption{Measurements of $\Afbzc$ at the Z pole.}
\label{fig:hf-bar-c}
\end{center}
\end{figure}

The combination has a rather low $\chi^2$ of 47.6 for $(105-14)$
degrees of freedom. The low $\chi^2$ is mainly caused by two effects:
for the rate measurements, several systematic uncertainties are
studied by varying parameters in the MC simulation or evaluated from
data.  While no effect is observed, the statistical accuracy of the
test is taken as a systematic uncertainty. In case of the asymmetries,
all measurements are very consistent, as shown in
Figures~\ref{fig:hf-bar-b} and~\ref{fig:hf-bar-c}, although they as
well as their combination are still statistics limited.

\begin{figure}[tbp]
\begin{center}
\vskip -0.75cm
\includegraphics[width=0.8\linewidth]{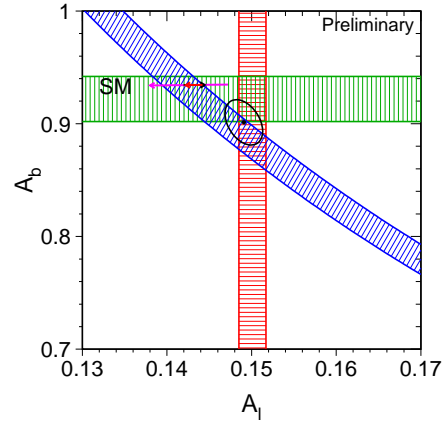}
\vskip -1.5cm
\caption{Bands of $\pm1$ standard deviation width showing the
  combined results of $\Al$, $\Ab$, and $\Afbzb$.  The SM expectation
  is shown as the arrows for $\MT=174.3\pm5.1~\GeV$ and
  $\MH=300^{+700}_{-186}~\GeV$.}
\label{fig:coup:aq}
\end{center}
\end{figure}

The mutual consistency of the measurements of $\Aq$,
$\Afbzq=(3/4)\Ae\Aq$ and $\Al$ assuming lepton universality is shown
in Figure~\ref{fig:coup:aq} for b quarks.  Compared to the
experimental uncertainties, the SM predictions are nearly constant in
$\Aq$, in contrast to the situation for $\Al$.  This is a consequence
of the SM values of the electric charge and of the iso-spin for
quarks.  


Since the leptonic asymmetry parameter $\Al$ is already well
determined, the measurements of $\Afbzq$ improve the determination of
the quark asymmetry parameters $\Aq$.  In the combined analysis, all
of the resulting asymmetry parameters $\Af$ are decreased in value
compared to their direct measurements, as shown by the contour curve
in Figure~\ref{fig:coup:aq}.  Compared to the SM expectation, the
combined extracted value for $\Ab$ is lower by 2.6 standard
deviations, while there is very good agreement for c quarks.

\subsection{Effective Electroweak Mixing Angle}

\begin{figure}[tbp]
\begin{center}
\includegraphics[width=0.8\linewidth]{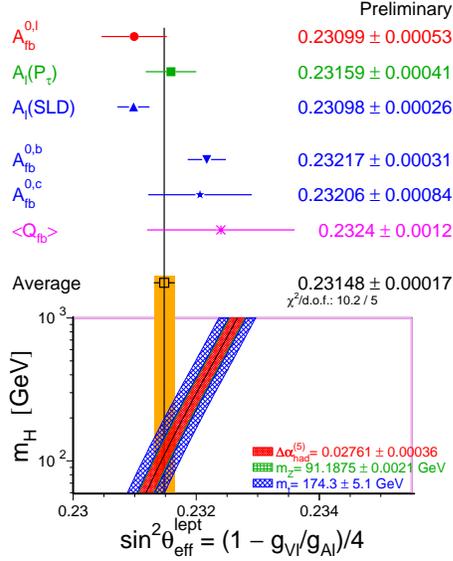}
\vskip -1cm
\caption{The effective electroweak mixing angle derived
  from various asymmetry measurements.}
\label{fig:sef2}
\end{center}
\end{figure}

Assuming the SM structure of the effective coupling constants, the
measurements of the various asymmetries are compared in terms of
$\swsqeffl$ in Figure~\ref{fig:sef2}.  The measurements fall into two
sets of three results each.  In the first set, the results on
$\swsqeffl$ are derived from measurements depending on leptonic
couplings only, $\Afbzl$, $\Al(P_\tau)$ and $\Al$(SLD).  In this case,
only lepton universality is assumed, and no further corrections to
interpret the results in terms of $\swsqeffl$ are necessary.  In the
second set, consisting of $\Afbzb$, $\Afbzc$ and the hadronic charge
asymmetry ${\langle Q_{\mathrm{FB}} \rangle}$, quark couplings are
involved.  In this case, the small non-universal flavour-specific
electroweak corrections, making $\swsqeffl$ different from that for
quarks, must be taken from the SM.  The effect of these corrections
and their uncertainties on the extracted value of $\swsqeffl$ is,
however, negligible.
The average of all six $\swsqeffl$ determinations is:
\begin{eqnarray}
\swsqeffl & = & 0.23148\pm0.00017\,,
\label{eq:coup:swsqeffl}
\end{eqnarray}
with a $\chi^2/dof$ of 10.2/5, corresponding to a probability of
7.0\%. The enlarged $\chi^2/dof$ is solely driven by the two most
precise determinations of $\swsqeffl$, namely those derived from the
measurements of $\Al$ by SLD, dominated by the left-right asymmetry
result, and of $\Afbzb$ at LEP.  These two measurements differ by 2.9
standard deviations.  Thus, the two sets of measurements, yielding
average values for $\swsqeffl$ of $0.23113\pm0.00021$
($\chi^2/dof=1.6/2$) and $0.23217\pm0.00029$ ($\chi^2/dof=0.05/2$),
respectively, also differ by 2.9 standard deviations.  This is a
consequence of the same effect as discussed before: the deviation in
$\Ab$ as extracted from $\Afbzb$ is reflected in the value of
$\swsqeffl$ extracted from $\Afbzb$.

\section{W AND FOUR-FERMION PRODUCTION}

\begin{figure}[tb]
\begin{center}
\includegraphics[width=0.8\linewidth]{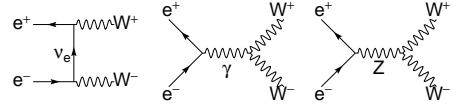}
\vskip -1cm
\caption{Lowest-order Feynman diagrams for $\PWW$ 
  production in $\Pee$ interactions. }
\label{fig:ww-prod}
\end{center}
\end{figure}

\begin{figure}[tb]
\begin{center}
\vskip -0.5cm
\includegraphics[width=0.8\linewidth]{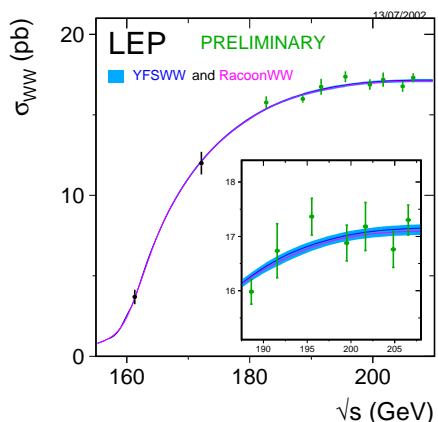}
\vskip -1.5cm
\caption{Measured $\PWW$ cross section compared to
  the SM expectations.}
\label{fig:ww-xsec}
\end{center}
\end{figure}

In recent years many measurements pertaining to the W boson have been
performed by the experiments taking data at LEP-2. The dominant
production of on-shell W bosons proceeds via the CC03 process of
W-pair production shown in Figure~\ref{fig:ww-prod}. The cross section
of this process is measured from the kinematic threshold threshold at
$\sqrt{s}=161~\GeV$ up to $\sqrt{s}=209~\GeV$~\cite{LEP2-WW}. The
preliminary combined LEP results are shown in
Figure~\ref{fig:ww-xsec}. Good agreement with the SM expectation is
observed, showing in particular the need for the triple-gauge-boson
vertices $\gamma$WW and ZWW.  The average of the ratio of measured to
predicted cross section has an accuracy of 1.1\%.  This accuracy
requires the calculation of ${\cal O}(\alpha)$ radiative corrections
in W-pair production~\cite{LEP2-TU}, which lower the cross section
prediction by about $(2.5\pm0.5)\%$.  Using the Monte Carlo event
generators YFSWW~\cite{YFSWW} (RacoonWW~\cite{RacoonWW}) containing
these corrections, a value for the ratio of 0.997 (0.999), in perfect
agreement with unity, is obtained.

\begin{figure}[tbp]
\begin{center}
\includegraphics[width=0.9\linewidth]{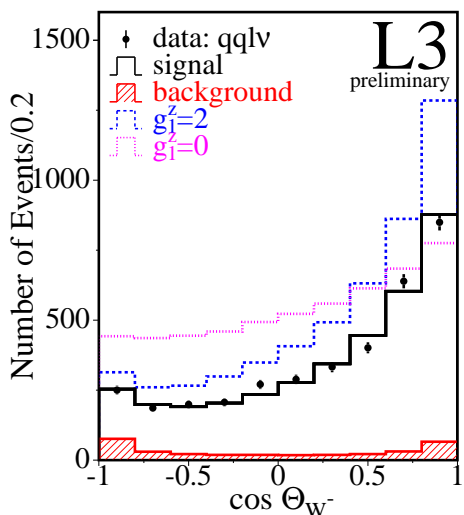}
\vskip -1.5cm
\caption{Polar scattering angle of the W$^-$ boson
  as observed by L3 in semileptonic W-pair events compared
  expectations in the SM $(g_1^Z=1)$ and for anomalous TGCs
  $(g_1^Z=0,2)$. }
\label{fig:ww-dxsec}
\end{center}
\end{figure}

The ${\cal O}(\alpha)$ corrections also affect the differential W-pair
cross section. The distribution in the polar scattering angle
$\cos\theta$ becomes steeper by about $2\%$. With nearly 10,000 W-pair
events per experiment, such a change is significant. In particular, it
mimics the effects of anomalous triple-gauge-boson couplings (TGCs),
which also modify the $\cos\theta$ distribution as shown in
Figure~\ref{fig:ww-dxsec}.  New TGC combinations from LEP-2 were
presented at this conference which for the first time take these
effects fully into account.

\begin{figure}[tbp]
\begin{center}
\includegraphics[width=0.8\linewidth]{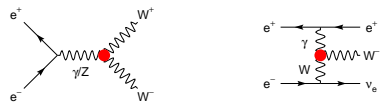}
\vskip -1cm
\caption{Some lowest order Feynman diagrams containing
  three-boson vertices. }
\label{fig:single-W}
\end{center}
\end{figure}

\begin{figure}[tbp]
\begin{center}
\vskip -0.5cm
\includegraphics[width=0.8\linewidth]{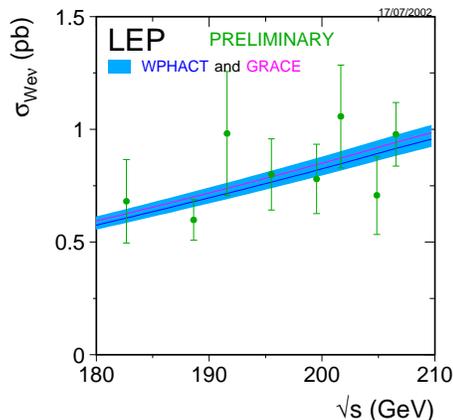}
\vskip -1.5cm
\caption{Measured cross section for single-W production compared to
  the SM expectation. }
\label{fig:1w-xsec}
\end{center}
\end{figure}

The TGCs considered are $g_1^Z$, $\kappa_\gamma$ and $\lambda_\gamma$,
where $g_1^Z$ is the weak charge of the W boson, \ie, it's coupling
strength to the Z, and $\kappa_\gamma$ and $\lambda_\gamma$ are
related to the magnetic dipole and electric quadrupole moment of the W
boson, $\mu_W = e(1+\lambda_\gamma+\kappa_\gamma)/(2\MW)$ and
$Q_W=e(\lambda_\gamma-\kappa_\gamma)/\MW^2$, respectively.  Besides
$\PWW$ production, additional constraints on the $\gamma$WW vertex
arise from single-W production, Figure~\ref{fig:single-W}.  The cross
section for this process, shown in Figure~\ref{fig:1w-xsec}, is
particularly sensitive to $\kappa_\gamma$.


The preliminary LEP-2 results are~\cite{LEP2-CTGC}:
\begin{eqnarray}
g_1^Z          & = & +0.998 \pm 0.024 \qquad\quad \hbox{(SM:~1)}\\
\kappa_\gamma  & = & +0.943 \pm 0.055 \qquad\quad \hbox{(SM:~1)}\\
\lambda_\gamma & = & -0.020 \pm 0.024 \qquad\quad \hbox{(SM:~0)}\,
\end{eqnarray}
where each of these TGCs is determined assuming all other couplings to
have their SM value. The uncertainty on the TGCs is dominated by the
theoretical uncertainty on the $\cos\theta$ slope.  Currently the full
${\cal O}(\alpha)$ correction to the slope is taken as theoretical
uncertainty, amounting to about 2/3 of the total error on the TGCs.
Studies are ongoing to evaluate the remaining theoretical uncertainty
on the $\cos\theta$ shape~\cite{LEP2-TU} and to propagate it to the
TGCs.


\begin{figure}[tbp]
\begin{center}
\includegraphics[width=0.4\linewidth]{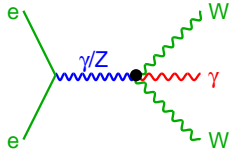}
\vskip -1cm
\caption{Lowest-order Feynman diagram in $\PWW\gamma$ production
  with a four-boson vertex. }
\label{fig:wwg-diag}
\end{center}
\end{figure}

\begin{figure}[tbp]
\begin{center}
\vskip -0.5cm
\includegraphics[width=0.7\linewidth]{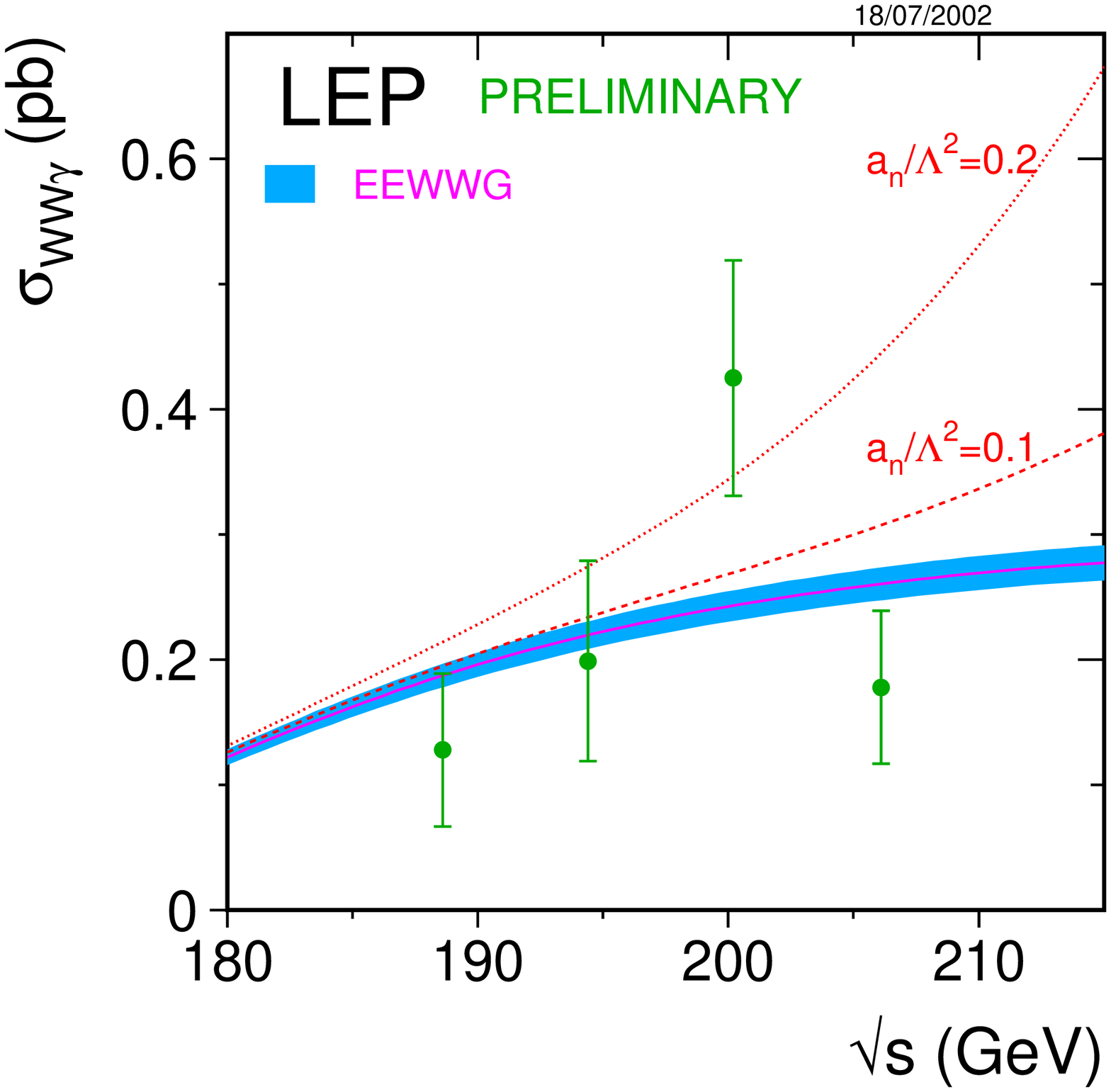}
\vskip -1cm
\includegraphics[width=0.7\linewidth]{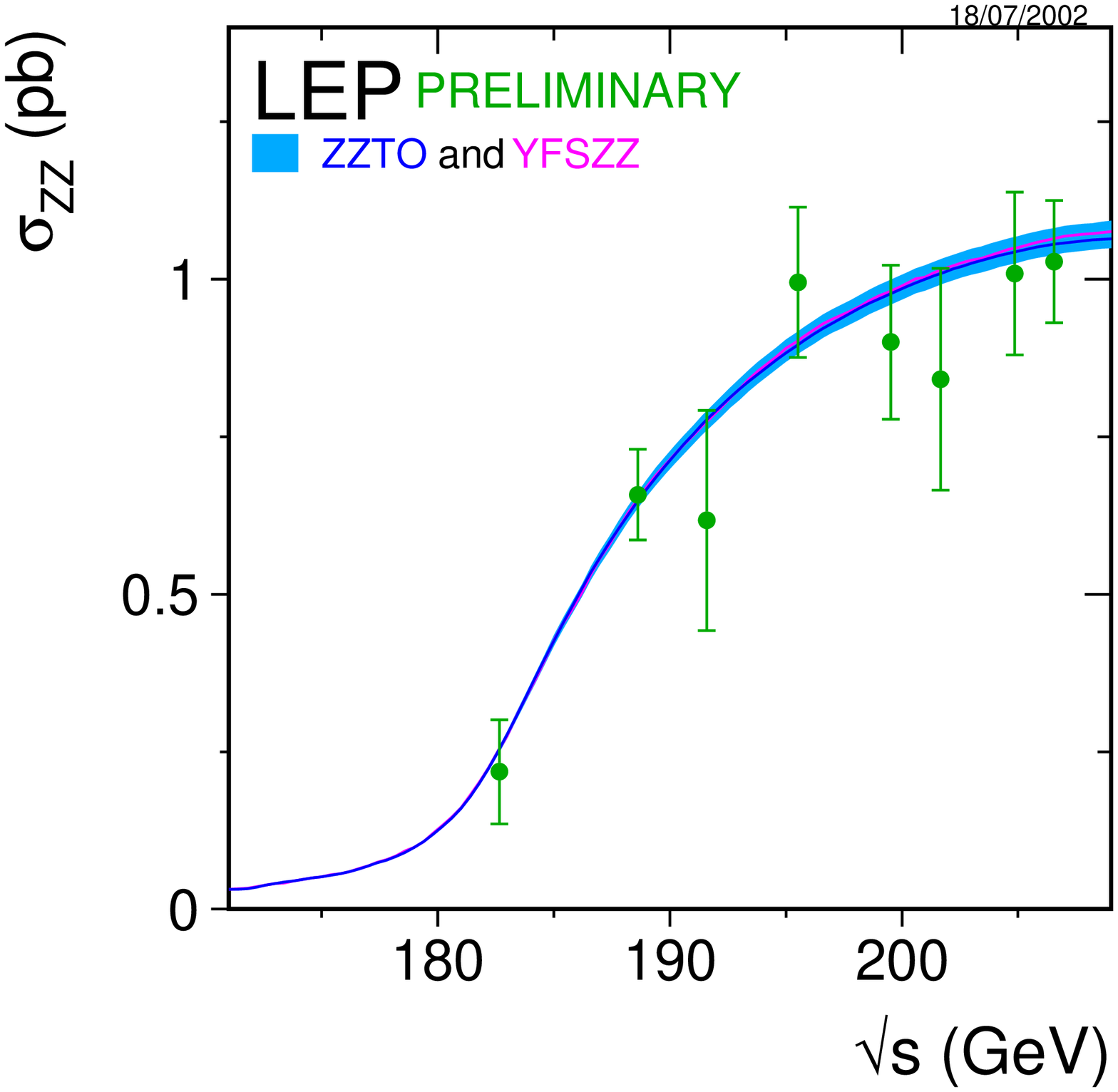}
\vskip -1cm
\includegraphics[width=0.7\linewidth]{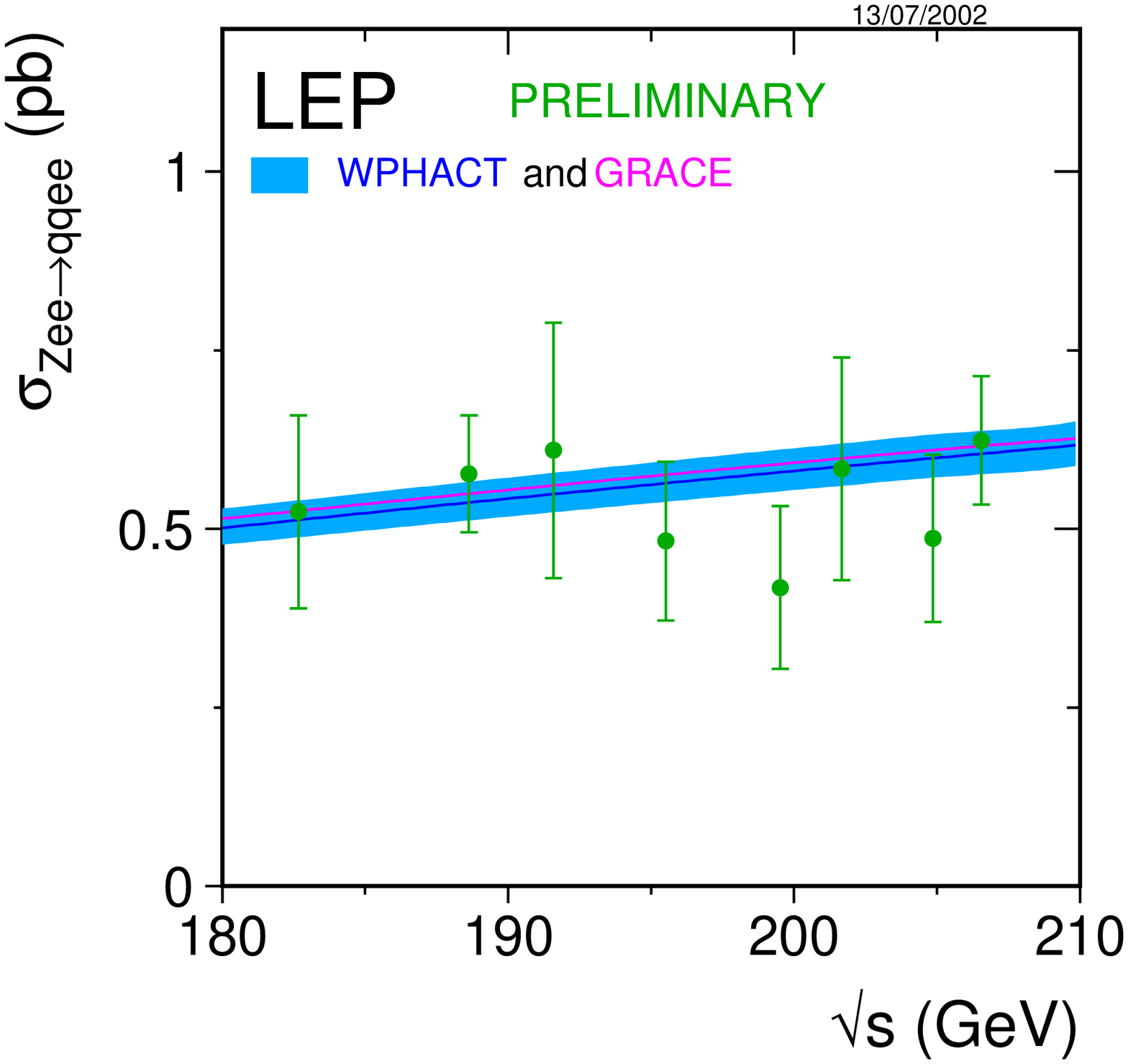}
\vskip -1.5cm
\caption{Measured cross sections for radiative W-pair production, ZZ
  and Zee production, compared to the SM expectations. }
\label{fig:small-xsec}
\end{center}
\end{figure}

Many more small-cross-section processes are now measured at
LEP-2~\cite{LEP2-WW,LEP2-4F}, for example radiative W-pair production
sensitive to quartic gauge boson couplings, Figure~\ref{fig:wwg-diag},
Z-pair production or Zee production.  As shown in
Figure~\ref{fig:small-xsec}, such small cross sections, a factor of up
to 100 smaller than that of W-pair production, are now measured with
an accuracy ranging from 5\% to 10\% and found to be in agreement with
the SM expectation.  These processes allow the determination of
neutral triple gauge couplings and of quartic gauge
couplings~\cite{LEP2-NQGC}.

\section{W-BOSON MASS AND WIDTH}

\begin{figure}[tbp]
\begin{center}
\includegraphics[width=0.8\linewidth]{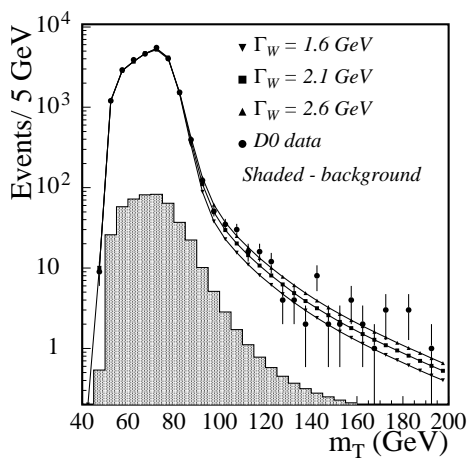}
\vskip -1cm
\caption{Transverse mass reconstructed in
  $W\rightarrow e \nu$ events by the D\O\ experiment in Run-I.  }
\label{fig:d0-mt}
\end{center}
\end{figure}

Until a few years ago, the W boson mass and width was measured at
hadron colliders only, most recently by the experiments CDF and D\O\ 
taking data at the Fermilab proton-antiproton TEVATRON collider.
Leptonic W decays with electrons and muons are selected and
reconstructed. The transverse mass, \ie, the invariant mass of the
lepton and the missing momentum vector in the plane transverse to the
beam axis is unaffected by the unknown longitudinal boost of the W
boson and bounded from above by the invariant mass of the decaying W
boson.  The distribution of the transverse mass is shown in
Figure~\ref{fig:d0-mt}.  The sharp upper edge, also called Jacobian
peak, yields the mass of the W boson, while the W-boson width is
derived from the high-mass tail of this distribution.  Final results
on $\MW$ and $\GW$ from CDF and D\O\ are now available for the
complete Run-1 data set.  They are combined taking correlations
properly into account~\cite{TEV-MW-GW}.

\begin{figure}[tbp]
\begin{center}
\vskip -0.5cm
\includegraphics[width=1.0\linewidth]{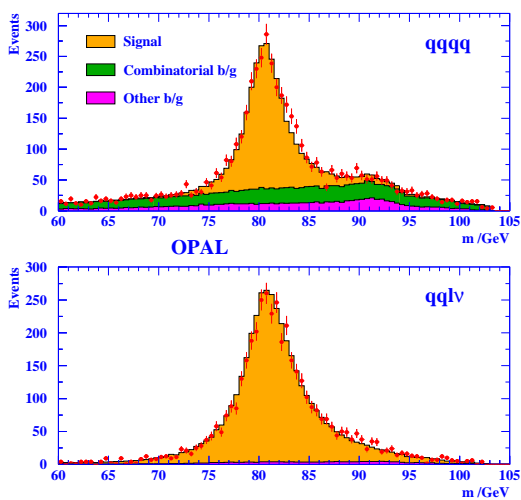}
\vskip -1cm
\caption{Invariant mass of W bosons
  reconstructed in (top) hadronic and (bottom) semileptonic W-pair
  events by the OPAL experiment. }
\label{fig:opal-mw}
\end{center}
\end{figure}

Since 1996 the W boson mass and width is also measured at LEP using
$\Pee\rightarrow\PWW\rightarrow\Pffff$ events.  Four-fermion final
states are selected and the two decaying W bosons are reconstructed.
For hadronic and semileptonic W-pair events, the W-pair kinematic is
completely reconstructed so that one directly measures the invariant
masses of the decaying W bosons as shown in
Figure~\ref{fig:opal-mw}~\cite{LEP2-MW-GW}, where the width of the
mass peak is given by detector resolution and the natural width of the
W boson.

\subsection{Final State Interconnection Effects}

\begin{figure}[tbp]
\begin{center}
\vskip -0.5cm
\includegraphics[width=0.5\linewidth]{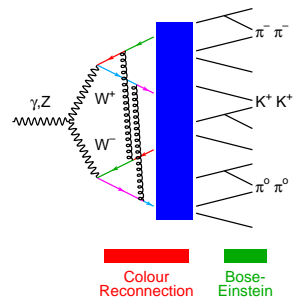}
\vskip -1.0cm
\caption{FSI effects: CR and BEC. }
\label{fig:ww-fsi-cartoon}
\end{center}
\end{figure}

For hadronic W-pair events, $\Pee\rightarrow\PWW\rightarrow\Pqqqq
\rightarrow hadrons$, cross talk effects may occur between the two
hadronic systems as shown in Figure~\ref{fig:ww-fsi-cartoon}.  The
four-momentum exchange causes the mass of the decaying W bosons to be
different from the mass of the hadronic decay products, thus leading
to potentially large systematic effects.

Such final-state interconnection (FSI) effects are colour reconnection
(CR), yielding a rearrangement of the colour flow in the perturbative
and non-perturbative phase of the parton shower evolution, and
Bose-Einstein correlations (BEC) between identical mesons in the final
state.  Recent analyses use the data to limit possible FSI bias on
$\MW(qqqq)$.

\begin{figure}[tbp]
\begin{center}
\includegraphics[width=0.8\linewidth]{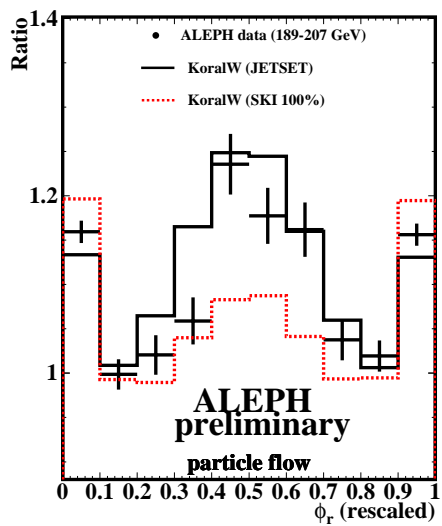}
\vskip -1cm
\caption{Ratio of particle flow distributions. }
\label{fig:pflow-aleph}
\end{center}
\end{figure}

CR~\cite{LEP2-CR} is searched for by studying the particle flow in the
two regions A and B between jets from the same W boson, versus that in
the two regions C and D between jets originating from different W
bosons, by projecting onto the planes defined by each of the regions A
to D.  In order to enhance the statistical sensitivity, the
distributions in the intra-W regions A and B are added and divided by
the sum of the inter-W regions C and D. The ratio R=(A+B)/(C+D) is
analysed as a function of the local angle between jets, as shown in
Figure~\ref{fig:pflow-aleph}. It can be seen that CR leads to a
depletion of R in the region between jets.

\begin{figure}[tbp]
\begin{center}
\includegraphics[width=0.8\linewidth]{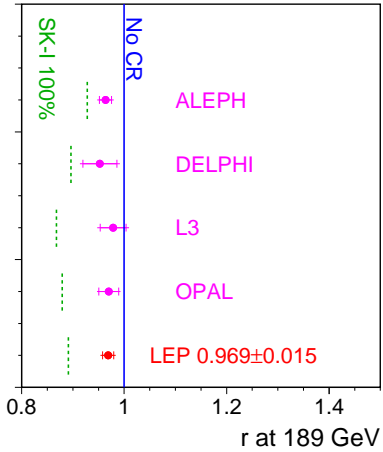}
\vskip -1cm
\caption{Ratio of the particle flow integrals as measured by the LEP
  experiments, compared to the expectations for no-CR and 100\% CR
  probability in the SK-I model. }
\label{fig:cr-lep}
\end{center}
\end{figure}

In order to quantify the measurement, the particle flow in A+B and in
C+D is integrated over the central region between jets and the ratio
is taken, yielding a single number.  The result obtained on data is
calibrated in sensitivity and central value to MC models incorporating
different models of CR. The results of the four LEP experiments are
shown in Figure~\ref{fig:cr-lep}.  Combining the results of the four
LEP experiments weighted by their sensitivity to a given CR model such
as SK-1, one observes a small hint, at the level of 2 standard
deviations off the non-CR scenario, for the presence of CR in the
data, corresponding to about 40\% CR in the SK-1 model.

For a given CR probability, all four experiments observe the same mass
shift in the SK-I model.  It is limited to be less than $90~\MeV$
averaged over centre-of-mass energies. All other CR models yield
smaller shifts. Since this preliminary result already includes most of
the full LEP-2 statistics, the limit on the bias will not improve
much.  The solution currently under investigation is thus to use jet
and event reconstruction algorithms less sensitive to CR effects, \eg,
to use only the core of the jets or to disregard low energy particles.

\begin{figure}[tbp]
\begin{center}
\includegraphics[clip=true,bb=35 60 640 525,width=0.8\linewidth]{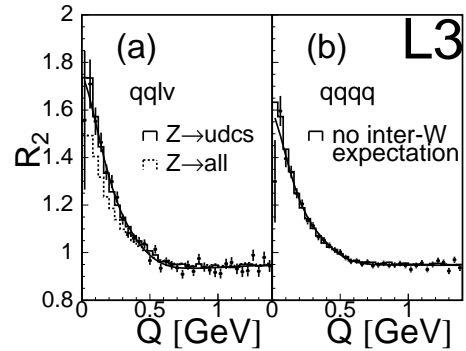}
\vskip -1cm
\caption{Rate of like-sign pairs normalised to the
  rate of unlike-sign pairs as a function of Q.  Left: Semileptonic
  W-pair events compared to Z decays; right: hadronic W-pair events. }
\label{fig:be-l3}
\end{center}
\end{figure}

BEC~\cite{LEP2-BE} between identical mesons in hadronic Z decays are
well established at LEP. They manifest themselves as an increased rate
of pairs of identical hadrons, or simply charged tracks in the
detector, which are close in phase space, \ie, at low four-momentum
difference Q.  Indeed the strength and shape of BEC observed in the Q
distribution is the same for hadronic W decays and hadronic Z decays
excluding b quarks as those do not arise in W decay, as shown in
Figure~\ref{fig:be-l3}.

\begin{figure}[tbp]
\begin{center}
\includegraphics[width=0.8\linewidth]{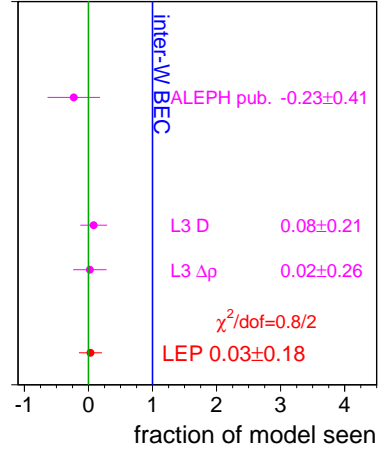}
\vskip -1cm
\caption{Measured fraction of inter-W BEC. }
\label{fig:be-lep}
\end{center}
\end{figure}

In order to be sensitive only to BEC between tracks from different W
bosons, the BEC effects from single W bosons are divided out in the
ratio D(Q).  Only inter-W BEC potentially affect the W-mass
reconstruction.  ALEPH and L3 do not find any sign of inter-W BEC,
with the combined result being that $(3\pm18)\%$ of the full effect of
inter-W BEC is observed, Figure~\ref{fig:be-lep}, in good agreement
with zero.  This corresponds to a bias on $\MW(qqqq)$ of less than
10~\MeV\ at 68\% CL, while the full effect would correspond to about
$35~\MeV$ for this particular model.

\begin{figure}[tbp]
\begin{center}
\includegraphics[clip=true,bb=0 255 460 490,width=0.9\linewidth]{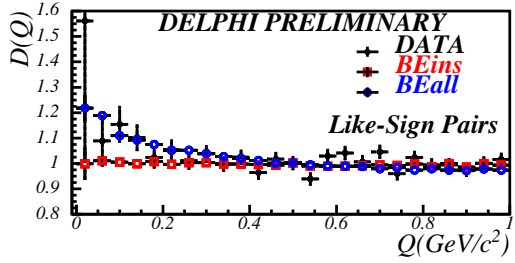}
\vskip -1cm
\caption{D(Q) sensitive to inter-W
  correlations only as measured by the DELPHI collaboration. }
\label{fig:be-delphi}
\end{center}
\end{figure}

However, the new preliminary DELPHI analysis observes an effect in the
D(Q) distribution as shown in Figure~\ref{fig:be-delphi}, with a
significance of 2.8 standard deviations away from zero.  Both results
differ by about two standard deviations. Studies are ongoing to
understand the different observations.

Based on the above studies, uncertainties of $90~\MeV$ for CR and
$35~\MeV$ for BEC are assigned to the mass of the W boson as extracted
from the four jet channel. Because of these large additional
uncertainties compared to the semileptonic channel, the weight of the
four jet channel in the LEP average is less than 10\%. The difference
in mass obtained for hadronic and semileptonic W-pair events,
calculated without FSI uncertainties, is $(9\pm44)~\MeV$, well
compatible with zero.

\subsection{Results}

\begin{figure}[tbp]
\begin{center}
\includegraphics[width=0.6\linewidth]{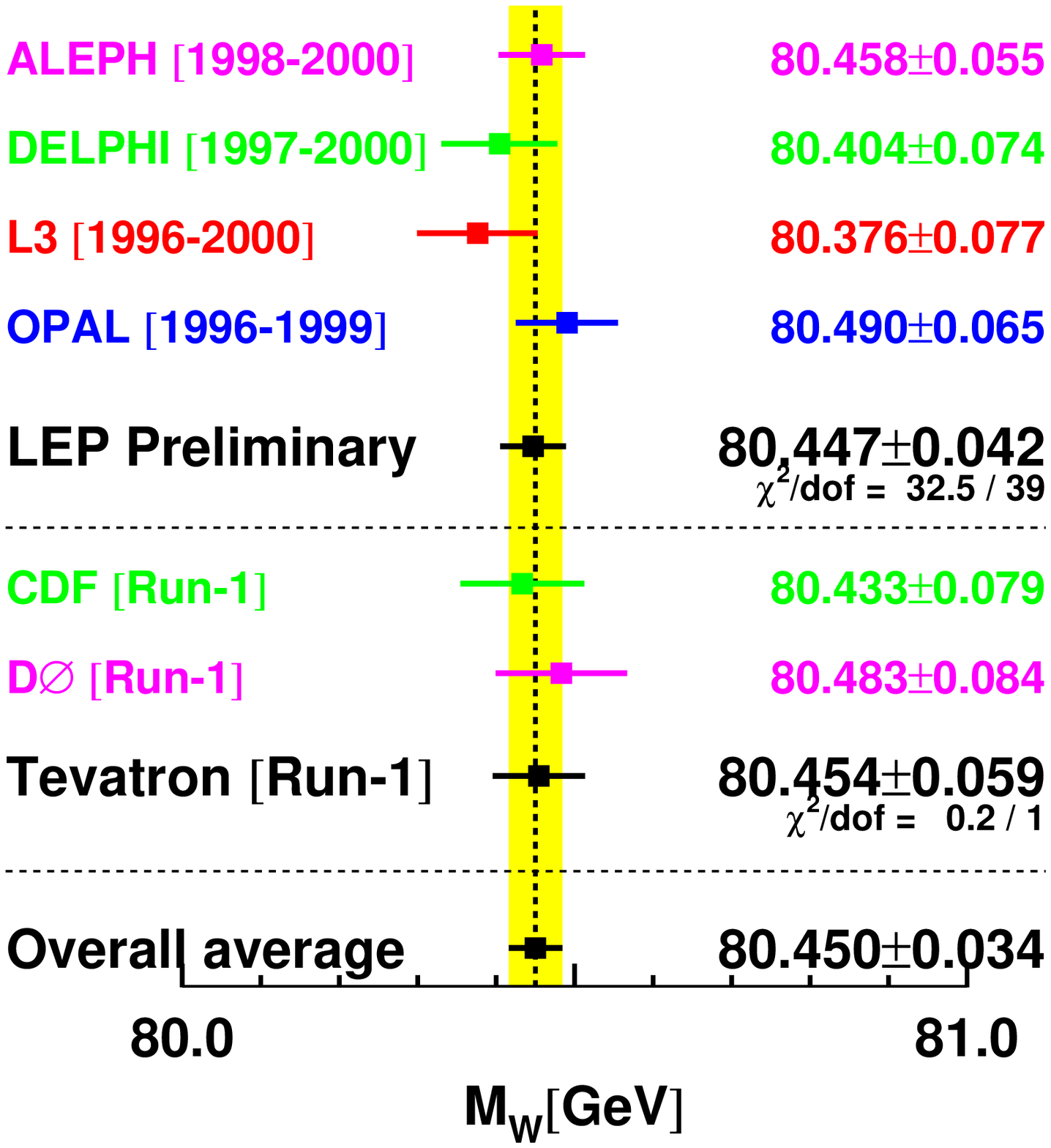}
\hskip -2cm
\includegraphics[width=0.6\linewidth]{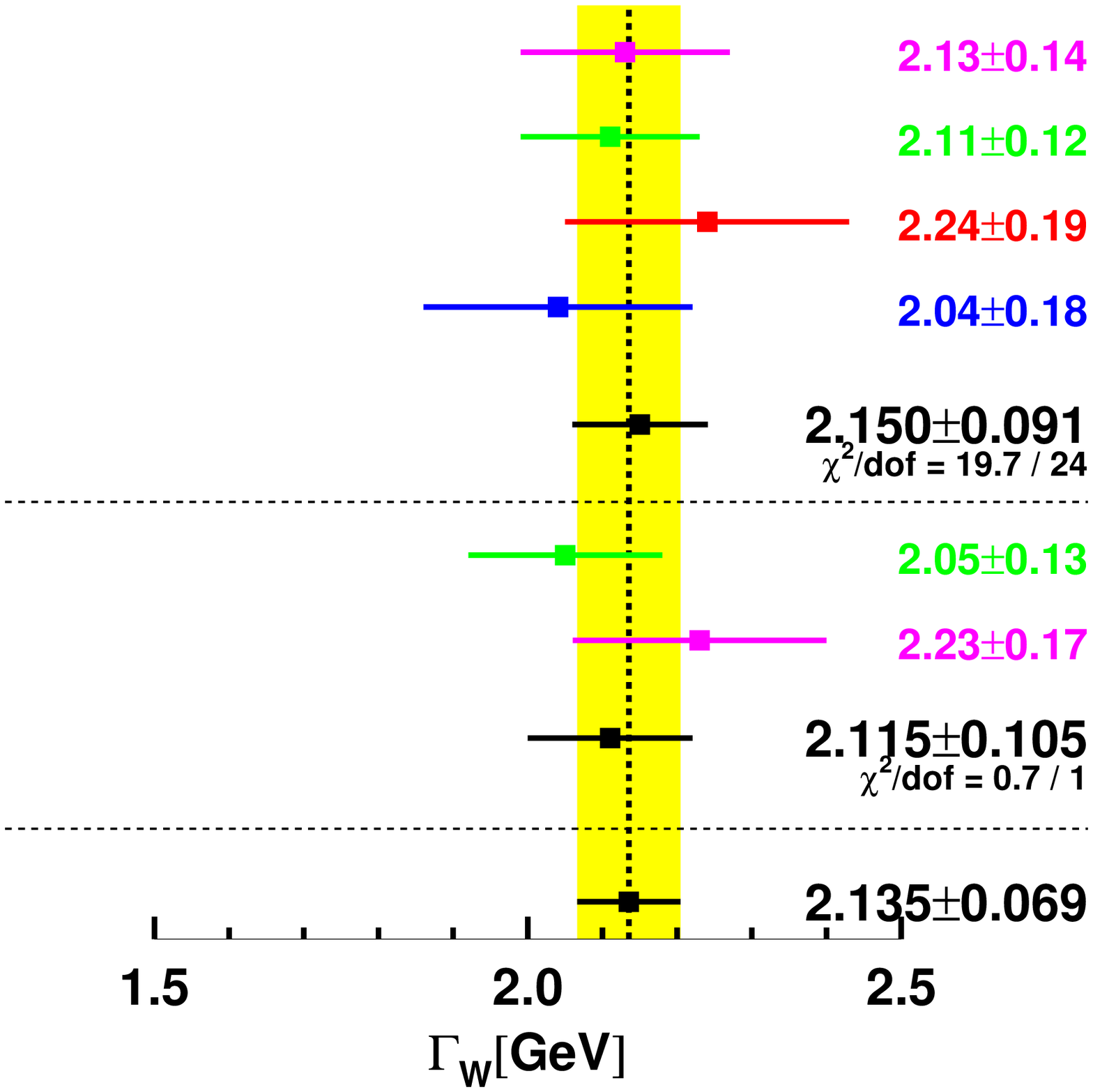}
\vskip -1cm
\caption{Results on $\MW$ and $\GW$ obtained from the
  six TEVATRON and LEP experiments. }
\label{fig:mw-gw-results}
\end{center}
\end{figure}

\begin{figure}[tbp]
\begin{center}
\vskip -0.75cm
\includegraphics[width=0.9\linewidth]{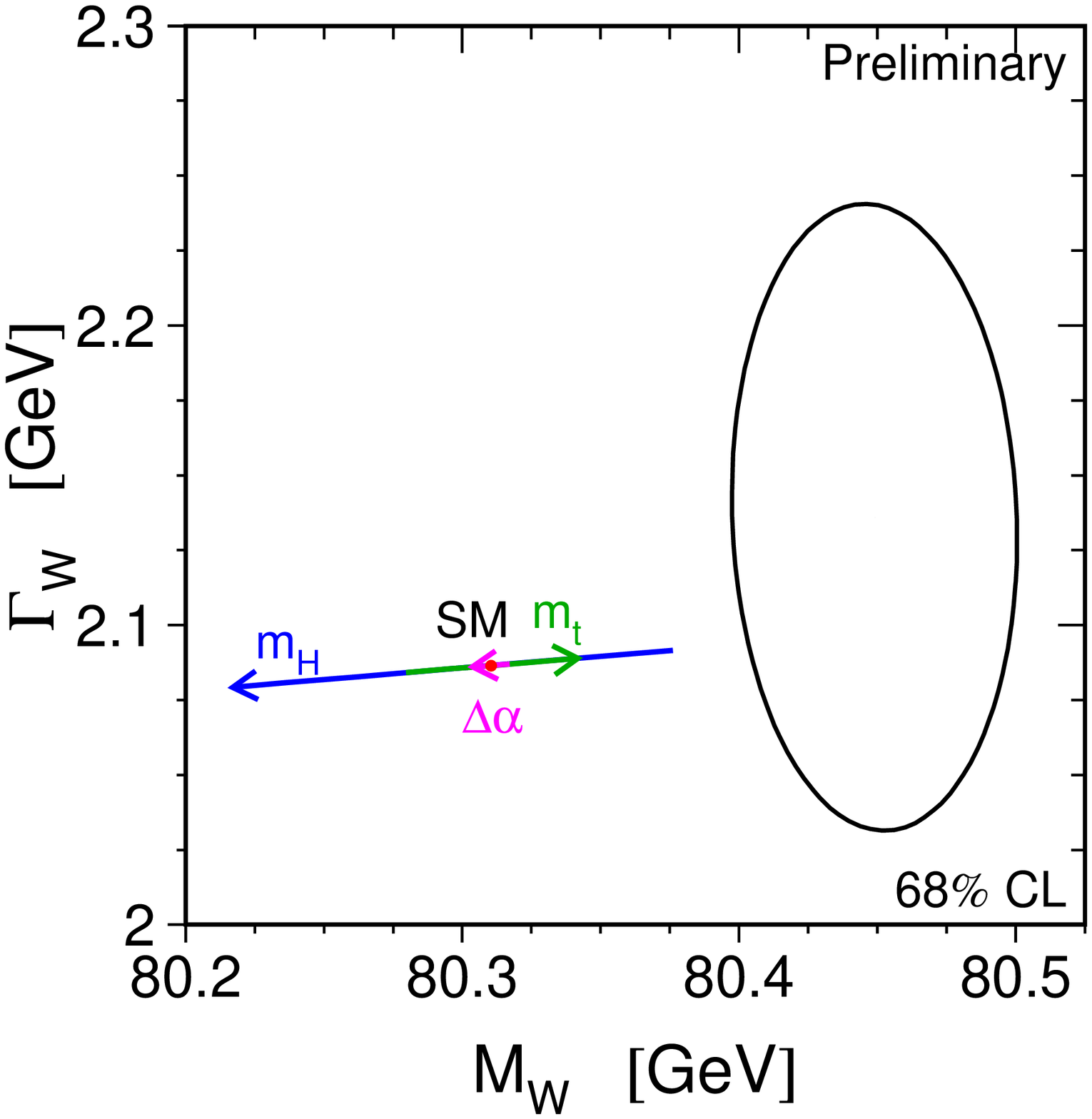}
\vskip -1.5cm
\caption{Contour curves of 68\% probability in the $(\MW,\GW)$ plane.
  The SM expectation is shown as the arrow for $\MT=174.3\pm5.1~\GeV$
  and $\MH=300^{+700}_{-186}~\GeV$.}
\label{fig:mw-gw-contour}
\end{center}
\end{figure}

The results of the six TEVATRON and LEP experiments on the mass and
the width of the W boson are shown in Figure~\ref{fig:mw-gw-results}.
Excellent agreement is observed.  The combined results and their
correlation is shown in Figure~\ref{fig:mw-gw-contour}, together with
the SM expectation.  It can be seen that the W mass is highly
sensitive to SM parameters, in particular preferring a low value for
the mass of the Higgs boson.


\section{TESTS OF THE STANDARD MODEL}

Within the framework of the SM, each of the observables presented
above are calculated as a function of five main relevant parameters,
which are the running electromagnetic and strong coupling constant
evaluated at the Z pole, $\dalhad$ and $\aqcd$, and the masses of Z
boson, top quark and Higgs boson, $\MZ$, $\MT$, $\MH$.  Using the
Fermi constant $\GF$ allows to calculate the mass of the W boson. The
running electromagnetic coupling is represented by the hadronic vacuum
polarisation $\dalhad$, as it is this contribution which has the
largest uncertainty, similar to the case of the muon anomalous
magnetic moment.

\begin{figure}[tbp]
\begin{center}
\includegraphics[width=0.3\linewidth]{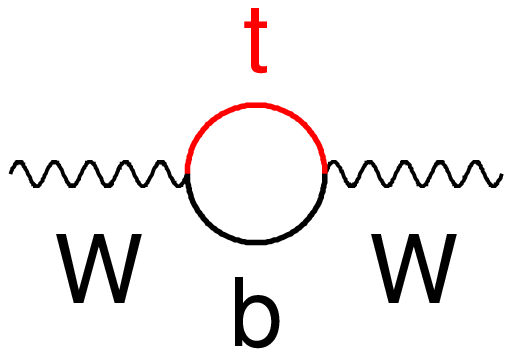}
\includegraphics[width=0.6\linewidth]{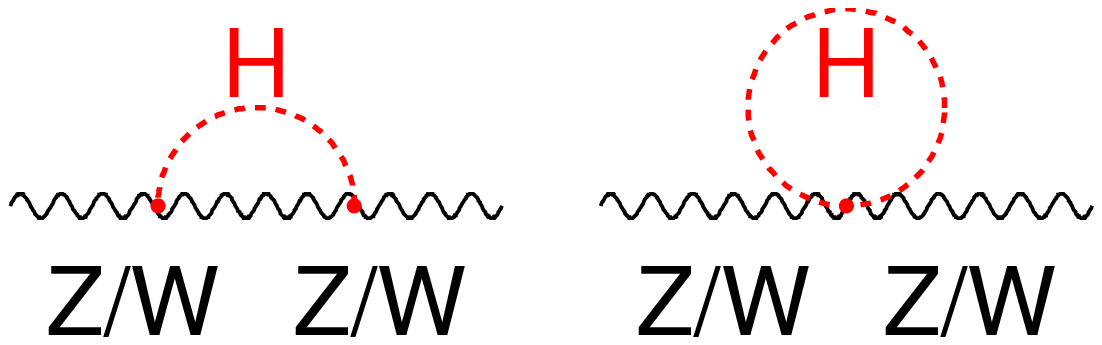}
\vskip -1cm
\caption{W and Z propagator corrections involving the top
  quark and the Higgs boson.}
\label{fig:ew-rad-cor}
\end{center}
\end{figure}

The precision of the Z-pole measurements require matching precision of
the theoretical calculations, first and second order electroweak and
QCD corrections etc. The dependence on $\MT$ and $\MH$ enters through
radiative corrections as depicted in Figure~\ref{fig:ew-rad-cor}.  The
computer programs TOPAZ0~\cite{TOPAZ0} and ZFITTER~\cite{ZFITTER},
incorporating state-of-the-art calculations, are used to calculate the
predictions as a function of the five SM parameters.  For the hadronic
vacuum polarisation, the result $\dalhad=0.02761\pm0.00036$~\cite{BOLEK},
based on the recent BES measurements~\cite{BES-R}, is used.

\begin{figure}[tbp]
\begin{center}
\vskip -0.75cm
\includegraphics[width=0.9\linewidth]{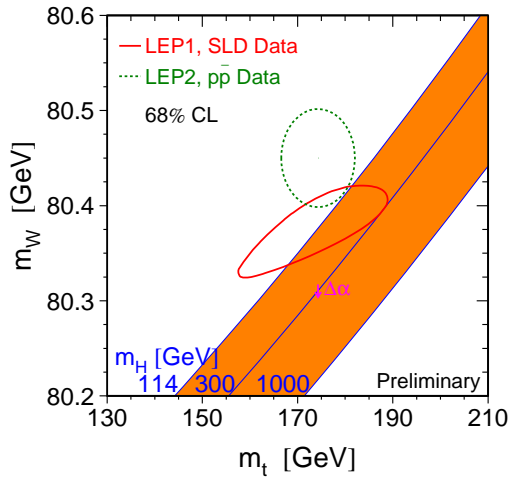}
\vskip -1.5cm
\caption{Contours of 68\% probability on the $(\MT,\MW)$ plane, for
  the corresponding direct and the indirect determinations. Also shown
  is the correlation between $\MW$ and $\MT$ as expected in the
  minimal SM for different Higgs boson masses. }
\label{fig:mt-mw-contour}
\end{center}
\end{figure}

Using the Z-pole measurements of SLD and LEP-1 in order to evaluate
electroweak radiative corrections, the masses of two heavy particles
measured at the TEVATRON and at LEP-2, namely the top quark and the
Higgs Boson, can be predicted. The resulting 68\% C.L. contour in the
$(\MT,\MW)$ plane is shown in Figure~\ref{fig:mt-mw-contour}. Also
shown is the contour corresponding to the direct measurements of both
quantities at the TEVATRON and at LEP-2. The two contours overlap,
successfully testing the SM at the level of electroweak radiative
corrections. The diagonal band in Figure~\ref{fig:mt-mw-contour} shows
the constraint between the two masses within the SM, which depends on
the mass of the Higgs boson, and to a small extend also on the
hadronic vacuum polarisation (small arrow labelled $\Delta\alpha$).
Both the direct and the indirect contour prefer a low value for the
mass of the SM Higgs boson.

\begin{figure}[tbp]
\begin{center}
\vskip -0.75cm
\includegraphics[width=0.9\linewidth]{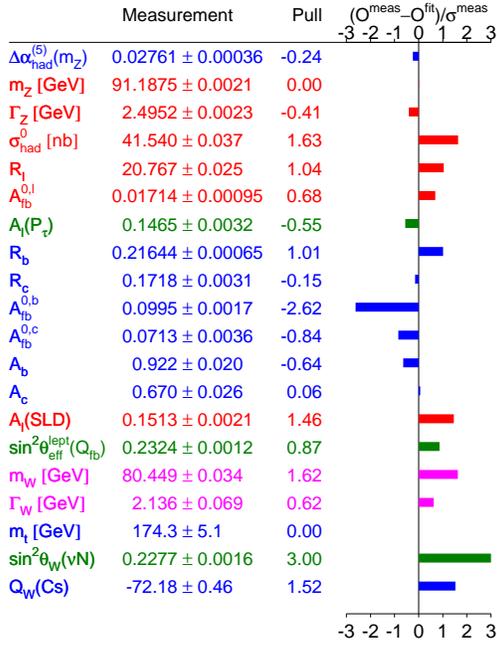}
\vskip -2cm
\caption{Pulls of all 20 measurements used in the global SM
  analysis. The pull is the difference between measured and expected
  value calculated for the minimum of the $\chi^2$, divided by the
  measurement error. }
\label{fig:pulls}
\end{center}
\end{figure}

The best constraint on $\MH$ is obtained by analysing all data.  This
joint fit has a $\chi^2$ of 29.7 for 15 degrees of freedom,
corresponding to a probability of only 1.3\%. The pulls of the 20
measurements entering the fit are shown in Figure~\ref{fig:pulls}.
The single largest contribution to the $\chi^2$, 9 units, arises from
the NuTeV measurement of the on-shell electroweak mixing angle.
Excluding the NuTeV measurement, the $\chi^2/dof$ becomes 20.5/14,
corresponding to 11.4\%. The fitted parameters in terms of central
value and error are almost unchanged, showing that the fit is robust
against the NuTeV result.

\begin{figure}[tbp]
\begin{center}
\vskip -0.75cm
\includegraphics[width=0.9\linewidth]{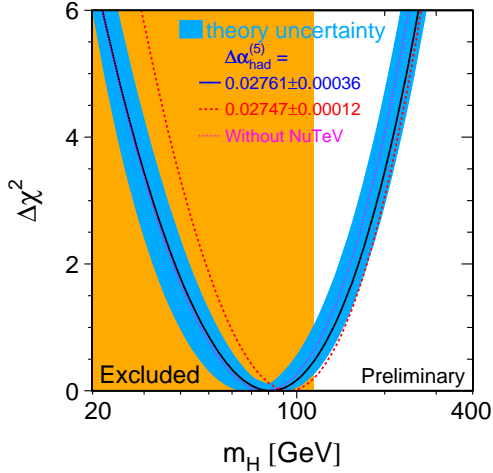}
\vskip -1.5cm
\caption{$\Delta\chi^2$ curve as a function of $\MH$.  
  Also shown are the curves using a theory-driven evaluation of the
  hadronic vacuum polarisation, or excluding the NuTeV measurement.}
\label{fig:sm-chi2}
\end{center}
\end{figure}

The global fit yields $\MH = 81^{+52}_{-33}~\GeV$, which corresponds
to a one-sided upper limit at 95\% C.L. on $\MH$ of $193~\GeV$
including theory uncertainty as shown in Figure~\ref{fig:sm-chi2}.
The fitted $\MH$ is strongly correlated with the hadronic vacuum
polarisation (correlation of -0.48) and the fitted top-quark mass
(0.70). The strong correlation with $\MT$ implies a shift of 35\% in
the predicted $\MH$ if the measurement of $\MT$ changes by one
standard deviation ($5~GeV$). Thus a precise experimental measurement
of $\MT$ is very important.

Also shown are the $\chi^2$ curves obtained with theory-driven and
thus more precise evaluation of the hadronic vacuum
polarisation~\cite{YNDURAIN}, or excluding the NuTeV result. Both
yield nearly the same upper limits on $\MH$. The theoretical
uncertainty on the SM calculations of the observables is visualised as
the thickness of the blue band. It is dominated by the theoretical
uncertainty in the calculation of the effective electroweak mixing
angle, for which a complete two-loop calculation is still lacking.

Also shown in Figure~\ref{fig:sm-chi2} is the $\MH$ range excluded by
the direct search for the Higgs boson, discussed in the following
section. Even though the minimum of the $\chi^2$ curve lies in the
excluded region, the uncertainties on the Higgs mass value are such as
that the results are well compatible.

\section{DIRECT HIGGS-BOSON SEARCH}

Only the direct observation of the Higgs boson constitutes a proof of
its existence; the indirect constraint on the Higgs boson mass already
assumes that the SM Higgs boson actually exists.  Direct searches for
the Higgs Boson have been carried out by the LEP experiments, with the
final results from LEP-2 now available~\cite{LEP2-Higgs}.

Higgs production at LEP occurs dominantly through radiation of a Higgs
Boson off an s-channel produced Z boson,
$\Pee\rightarrow\PZ\PH\rightarrow\Pff\Pbb$, yielding a four-fermion
final state. Smaller contributions to the cross section arise through
WW and ZZ fusion especially for Higgs masses beyond the nominal
kinematic limit of ZH production.

\begin{figure}[tbp]
\begin{center}
\vskip -0.5cm
\includegraphics[width=0.7\linewidth]{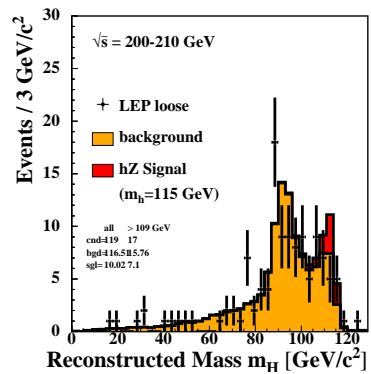}
\includegraphics[width=0.7\linewidth]{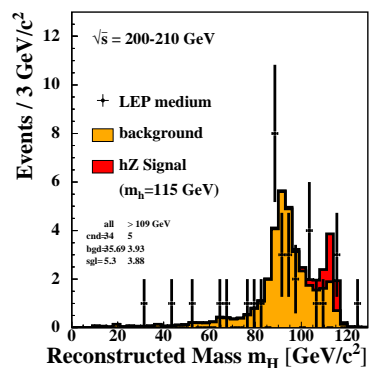}
\includegraphics[width=0.7\linewidth]{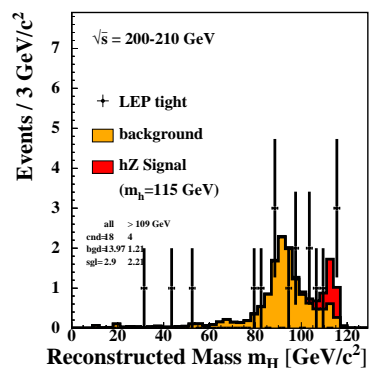}
\vskip -1cm
\caption{Reconstructed Higgs mass in ZH candidates
  selected by the four LEP experiments, for increasing purity of the
  selection. }
\label{fig:smhiggs-mass}
\end{center}
\end{figure}

The analysis combines Higgs-mass independent selections with
Higgs-mass reconstruction. For increasingly pure selections, the
distribution of reconstructed candidate Higgs boson masses is shown in
Figure~\ref{fig:smhiggs-mass}, with a small accumulation of events at
about $115~\GeV$.

\begin{figure}[tbp]
\begin{center}
\includegraphics[width=0.9\linewidth]{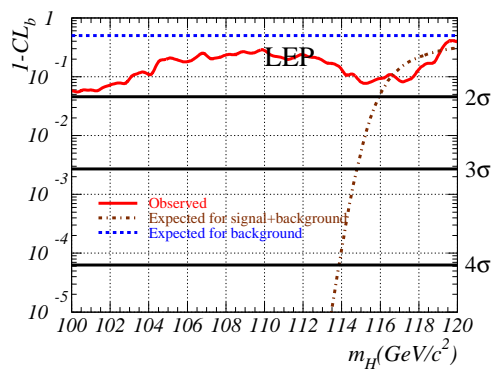}
\vskip -1cm
\caption{Confidence level for the background-only hypothesis.}
\label{fig:smhiggs-clb}
\end{center}
\end{figure}

The complete statistical analysis for the direct search is based on a
global discriminating variable, which combines all event information
and separates between signal and background events, and the
reconstructed Higgs mass. Figure~\ref{fig:smhiggs-clb} shows the
confidence level for the hypothesis of the observed data being SM
background only, as a function of hypothetical Higgs masses. The
largest excess over background at high masses shows a significance of
1.7 standard deviations, corresponding to a probability for a
background fluctuation of 8\%, at masses around $115$ to $117~\GeV$.
This small excess is solely given by one experiment, ALEPH, where the
significance is about 2.8-3.0 standard deviations, and in one channel
only, namely qqbb.

Without evidence for direct Higgs production in the LEP combined data
set, the data are used to set a lower limit on the mass of the SM
Higgs boson of $114.4~\GeV$ at 95\% C.L.
Because of the small excess, the observed limit is $0.9~\GeV$ smaller
than the expected limit.

\section{OUTLOOK}

Since 2001, Run-II of data taking at the TEVATRON is ongoing.  The
proton-antiproton centre-of mass energy is raised from $1.8~\TeV$ to
$1.96~\TeV$. Each experiment will collect about 15/fb in total, and
about 2/fb up to 2004~\cite{TEV-TOP,TEV-RunII}. Among other things,
this will allow measurements of $\MW$ with an accuracy of $25~\MeV$
and of $\MT$ with an accuracy of $2.5~\GeV$, yielding a Higgs mass
prediction based on electroweak radiative corrections of $30\%$
accuracy.  Besides the width of top quark and W boson, also gauge
couplings will be measured with high precision.  An interesting aspect
concerns Z production with subsequent decay into lepton pairs.  This
process allows to reconstruct the centre-of-mass system and thus the
forward-backward asymmetry in this system, which is a measurement of
the effective electroweak mixing angle.
The estimated precision is in the range of current Z-pole
determinations of $\swsqeffl$, thus would be useful to shed light on
the dispersion of the various existing measurements of that quantity.
Crucial for this measurement are precise matrix elements and parton
distribution functions including theoretical uncertainties.
Constraints on these could also be determined from W production data.


The TEVATRON experiments will perform the next step in the direct
search for the Higgs boson. With 15/fb per experiment, the two
experiments will be sensitive to Higgs masses of up to $135~\GeV$ at
the level of three standard deviations,
well beyond the existing limit.


\section{CONCLUSION}

During the last decade many experiments, such as BES, E821, NuTeV,
SLD, ALEPH, DELPHI, L3, OPAL, CDF and D\O, etc. have performed a
wealth of measurements of unprecedented precision in particle
physics. These measurements test all aspects of the SM of particle
physics, and many of them show large sensitivity to electroweak
radiative corrections at loop level.

Most measurements agree with the expectations as calculated within the
framework of the SM, successfully testing the SM at Born and at loop
level. Still there are two ``3 standard deviations effects'', namely
the spread in the various determinations of the effective electroweak
mixing angle, and NuTeV's result, most pronounced when interpreted in
terms of the on-shell electroweak mixing angle.

For the experiments at current and future colliders, precise
theoretical calculations are needed, which must include an assessment
of the remaining theoretical
uncertainty~\cite{LEP2-TU,CEEX,MultiLoop,Sudakov}.  The required
precision must match the expected experimental precision in
measurements of observables such at the masses of top quark and W
boson, the effective electroweak mixing angle and the hadronic vacuum
polarisation based on low-energy data.

The next qualitative step is surely given by the observation of the SM
Higgs boson or a Higgs boson of any other model.  Searches will be
performed at TEVATRON, while experiments at future linear
colliders~\cite{LC-Higgs} and notably the LHC~\cite{LHC-Higgs} will be
sufficiently powerful to find the Higgs boson or indications of other
mechanisms of electroweak symmetry breaking.

\section{ACKNOWLEDGEMENTS}

It is a pleasure to thank my colleagues of the TEVATRON and LEP
electroweak working groups, members of the BES, E821, NuTeV, SLD,
ALEPH, DELPHI, L3, OPAL, CDF and D\O\ experiments, as well as
D. Bardin, P.  Gambino, G. Passarino and G. Weiglein for valuable
discussions.

\end{document}